\documentclass[11pt]{article}
\pdfoutput=1
\usepackage{amsmath}
\usepackage{amssymb}
\usepackage{graphicx,bbm,mathrsfs}
\usepackage{nicefrac}
\usepackage{slashed}
\usepackage{bbm}
\usepackage{geometry}
\usepackage{empheq}
\usepackage{stackrel}
\usepackage{ulem}
\usepackage{xcolor}
\usepackage{physics}
\usepackage{placeins}
 
\usepackage{tocloft}

\usepackage{jheppub}
\usepackage{setspace}
\usepackage{relsize}

\usepackage{empheq}

\usepackage{pifont}

\usepackage{mathtools}
\usepackage{enumitem}
\usepackage{dcolumn}   
\usepackage{bm}        
\usepackage{graphicx,mathrsfs}
\usepackage{nicefrac}
\usepackage{multirow}
\usepackage{color}
\usepackage{mathtools}
\usepackage{makecell}

\usepackage{environ} 
\usepackage{lipsum} 
 \NewEnviron{Smaller11}{
           \scalebox{1.1}{$\BODY$} 
 } 
 \NewEnviron{Smaller08}{
           \scalebox{0.8}{$\BODY$} 
 }

\newcommand{\be}{\begin{equation}}
\newcommand{\ee}{\end{equation}}
\newcommand{\bea}{\begin{eqnarray}}
\newcommand{\eea}{\end{eqnarray}}

\newcommand{\h}{c}

\usepackage{wasysym}
\usepackage{hhline,colortbl}

\renewcommand{\h}{h}

\usepackage{titlesec}

\titleformat*{\section}{\Large\bfseries}
\titleformat*{\subsection}{\large\bfseries}
\titleformat*{\subsubsection}{\large\bfseries}
\titleformat*{\paragraph}{\large\bfseries}
\titleformat*{\subparagraph}{\large\bfseries}

\makeatletter
\newcommand*{\prodsym}{%
  \DOTSB
  \mathop{
    \mathchoice
      {\rlap{\kern.3em\rotatebox[origin=c]{-90}{}}{\prod}}
      {\vcenter{\rlap{\kern.2em\rotatebox[origin=c]{-90}{}}}{\prod}}
      {\sum}{\sum}
  }\slimits@
}
\makeatother

\makeatletter
\DeclareFontFamily{OMX}{MnSymbolE}{}
\DeclareSymbolFont{MnLargeSymbols}{OMX}{MnSymbolE}{m}{n}
\SetSymbolFont{MnLargeSymbols}{bold}{OMX}{MnSymbolE}{b}{n}
\DeclareFontShape{OMX}{MnSymbolE}{m}{n}{
    <-6>  MnSymbolE5
   <6-7>  MnSymbolE6
   <7-8>  MnSymbolE7
   <8-9>  MnSymbolE8
   <9-10> MnSymbolE9
  <10-12> MnSymbolE10
  <12->   MnSymbolE12
}{}
\DeclareFontShape{OMX}{MnSymbolE}{b}{n}{
    <-6>  MnSymbolE-Bold5
   <6-7>  MnSymbolE-Bold6
   <7-8>  MnSymbolE-Bold7
   <8-9>  MnSymbolE-Bold8
   <9-10> MnSymbolE-Bold9
  <10-12> MnSymbolE-Bold10
  <12->   MnSymbolE-Bold12
}{}

\let\llangle\@undefined
\let\rrangle\@undefined
\DeclareMathDelimiter{\llangle}{\mathopen}%
                     {MnLargeSymbols}{'164}{MnLargeSymbols}{'164}
\DeclareMathDelimiter{\rrangle}{\mathclose}%
                     {MnLargeSymbols}{'171}{MnLargeSymbols}{'171}
\makeatother

\begin{document}

\vspace*{4mm}

\thispagestyle{empty}

\begin{center}

\begin{minipage}{20cm}
\begin{center}
\hspace{-5cm }
\Huge
\sc
 Background-Induced Forces  \\
\hspace{-5cm }
from Dark Relics
\end{center}
\end{minipage}
\\[30mm]

\renewcommand{\thefootnote}{\fnsymbol{footnote}}

{\large  
Sergio~Barbosa\,\footnote{sergio.barbosa@aluno.ufabc.edu.br}, 
Sylvain~Fichet\,\footnote{sylvain.fichet@gmail.com}
}\\[12mm]
\end{center} 
\noindent

\textit{CCNH-Universidade Federal do ABC, Santo Andre, 09210-580 SP, Brazil}

\addtocounter{footnote}{-1}

\vspace*{15mm}
 
\begin{center}
{  \bf  Abstract }
\end{center}
\begin{minipage}{15cm}
\setstretch{0.95}

Light particles quadratically coupled to nucleons induce macroscopic forces in matter. 
While a quantum effect always exists, 
an additional force occurs in the presence of a finite density of the light particles. 
We compute and classify such background-induced forces for particles of spin $0,\frac{1}{2},1$ in the framework of effective field theory. 
We show that, at short distance, the background-induced forces exhibit a universal behavior that depends solely on the moments of the phase space distribution function of the light particles.

We compute the forces in the case of dark particles densities that may realistically occur in cosmology, assuming either \textit{(i)} cosmically homogeneous or \textit{(ii)} virialized phase space distributions. For homogeneous distributions --- analogous to cosmic neutrinos, all the background-induced forces remain, unlike the quantum ones, exponentially unsuppressed at large distance, implying that large scale fifth force experiments are highly sensitive to dark relics. Moreover at zero mass the forces from dark bosons are generically enhanced with respect to their quantum counterpart due to Bose-Einstein distribution. 
Overall, we find that the resulting fifth force bounds can compete with those from quantum forces.
For virialized distributions --- identifiable as cold dark matter, the reach is also enhanced  beyond the dark matter Compton wavelength. We obtain significant bounds on sub-keV scalar cold dark matter, that can appear in certain cosmological scenarios. A thorough adaptation of the results from the E\"ot-Wash experiment may produce powerful additional bounds.

    \vspace{0.5cm}
\end{minipage}

\newpage
\setcounter{tocdepth}{2}


\tableofcontents

\newpage

\clearpage

\section{Introduction \label{se:intro}}

The existence of a light particle quadratically coupled to nucleons results in the generation of forces between macroscopic bodies of ordinary matter.  There are two distinct versions of this phenomenon: \textit{i)} a  quantum, Casimir-Polder-like force induced from the exchange of virtual light particles,  and \textit{ii)} a background-dependent force induced in the presence of a finite density of real light particles. 
This work is dedicated to the latter phenomenon, that we refer to as background-induced forces.

The existence of light particles coupled via bilinear operators to Standard Model (SM) operators is a natural possiblity occurring in hypothetical extensions of the SM. Such particles can be part of a hidden sector  beyond the SM, that can also be referred to as \textit{dark} sector since it  interacts only weakly or not at all with photons.  
In the context of physics beyond the SM, the macroscopic forces \textit{i)} and \textit{ii)} can be used as probes of the dark sector.   We refer then to them as {\it dark forces}.
 
Strikingly, the background-induced force is not a mere correction to the quantum force, it behaves differently and can dominate over its quantum counterpart  in certain situations.  
An analogous phenomenon that occurs in condensed matter  is the Casimir force induced by thermal fluctuations in thermodynamic systems, which dominates over the quantum Casimir force when the system is at criticality \cite{Dantchev:2022hvy}.
The resulting thermal (or critical) Casimir force typically tends to have longer range than the quantum one, a feature that will also appear throughout our study.

In the particle physics context, the background-induced dark force may provide a way to search for  cosmological relics of dark particles.
We focus on some of these {\it dark relics} as a natural example of application. A sizeable part of the paper is, however, independent of  specific choices of the phase space distribution function.

The phenomenon of a background-induced force has first been pointed out and discussed in the context of neutrino physics in \cite{Horowitz:1993kw,Ferrer:1998ju,Ferrer:1999ad}.\,\footnote{
The quantum component of the neutrino-induced force has been gradually derived and studied in \cite{Feinberg:1968zz,Hsu:1992tg,Grifols:1996fk,Lusignoli:2010gw,Costantino:2020bei,LeThien:2019lxh, Segarra:2020rah, Xu:2021daf, Coy:2022cpt}.  A distinct neutrino-related effect occurs  in  forces generated by  dark particles linearly coupled to neutrinos, that experience screening  due to the effective mass induced by the  cosmic neutrino background \cite{Chauhan:2024qew}. }
Recently, the topic enjoyed renewed interest.
Ref.\,\cite{Ghosh:2022nzo}, with further clarifications provided in  
 \cite{Blas:2022ovz}, proposed experimental methods to 
probe neutrino physics via the neutrino-background induced force. 
Ref.\,\cite{Arvanitaki:2022oby} revisited the local cosmic neutrino background. 
{
A useful classical understanding of these forces has been developed in \cite{Hees:2018fpg} (see also \cite{Banerjee:2022sqg}) then \cite{VanTilburg:2024tst}, in which they are dubbed ``wake forces'' and some dark particles are investigated. }
Inspired by these works, we pursue our study of macroscopic dark forces induced by hidden sectors developed in \cite{Fichet:2017bng,Brax:2017xho,Brax:2018zfb,Costantino:2019ixl,Brax:2022wrt}.
As in our previous studies  we work within an effective field theory (EFT) framework with dark particles of spin $0,\frac{1}{2},1$, either self-conjugate or not.

The steps of our investigation are as follows. Section~\ref{se:propagators}  presents the propagators for scalar, spinor and vector fields in the presence of a finite density.
In section~\ref{se:EFT_Potentials} we define the EFT  and present the general steps of the background-induced force computations. We also correct a previously computed quantum potential from a spin-$1$ loop in which a Goldstone loop had been omitted. 
In section~\ref{se:ShortDistance} we present general properties of the background-induced potentials in the short distance limit and a relation to the moments of the phase space distribution function. 
{In section~\ref{se:BackgroundForcesRelics} and \ref{se:BackgroundForcesDarkMatter} we introduce  the homogeneous and virialized distributions motivated by standard cosmic history,  and  compute the potentials induced by dark relics with such distributions.}   In section~\ref{se:5thForces} we review experimental bounds and apply them to our potentials, hence bounding the parameter space of dark relics. 
{The case of a typical anisotropic distribution is explored in section \ref{sec: anisotropy}. 
}
Section~\ref{se:conclusion} contains our conclusion. The Appendix contains
a  derivation of the propagators (\ref{app:propagators}), details on the $V_b^1$ quantum potential (\ref{app:V1b}), the background-induced amplitudes (\ref{app:amplitudes}) and potentials (\ref{app:potentials}), and an example of derivation (\ref{app:example}).

\section{Propagators at Finite Density}
\label{se:propagators}

We present the Feynman propagators for scalar, fermion and vector fields. All  propagators take the form
\be
D(p)=
 \vcenter{
        \hbox{
        \includegraphics[trim={11cm 9cm 7cm 9cm},clip,width=0.3\textwidth]{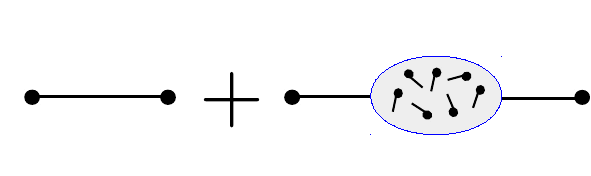}
        }
        } 
  \, \label{eq:prop_finite_density}
\ee
where the blob represents the contribution from the nonzero phase space distribution function of the real particles. 
The derivations in canonical quantization are presented in Appendix \ref{app:propagators}, see  also \cite{Ghosh:2022nzo, Blas:2022ovz,VanTilburg:2024tst}.

The key point is that, unlike in vacuum,  normal ordered products of creation and annihilation operators do not vanish when evaluated over the background.  
 Instead, the expectation value of a normal ordered product defines the associated phase space distribution of the background: 
\be
\langle a^\dagger_{\bm p} a_{\bm k} \rangle_{\rm background} = (2\pi)^3 \delta^{(3)}({\bm p}- {\bm k}) f({\bm k})\,.  
\ee
This leads to the second term in \eqref{eq:prop_finite_density}. 
A similar definition applies for antiparticles and to independent polarizations in case of particles with spin.

For a complex scalar field we find
\begin{equation}
    D_0(p) = \frac{i}{p^2 - m^2 + i\epsilon} + 2\pi \delta(p^2 - m^2)(\theta(p^0)f_{+}(\boldsymbol{p}) + \theta(-p^0)f_{-}(\boldsymbol{p})).
    \label{eq:Dscalar}
\end{equation}
The first term is the usual vacuum term interpreted  as the exchange of a virtual particle. The second term can be interpreted as an absorption of the on-shell particle into the background, followed by the emission of an on-shell particle with same $3$-momentum.

In case of fields with spin, the density may in general be polarized. This is encoded by having distinct phase space distribution functions  $f^s(\boldsymbol{p})$
in a given polarization basis labelled by $s$.  
 The unpolarized case is recovered when all distributions are equal,  $f^s(\boldsymbol{p})\equiv f(\boldsymbol{p})$ for all $s$.

For the fermion field the polarized and unpolarized propagators at finite density are respectively 
\begin{equation}
    D_{\frac{1}{2}}(p)|_{\rm pol.}  =  \frac{i (\slashed{p} + m)}{p^2 - m^2 + i\epsilon} - (2\pi) \delta(p^2 - m^2)\sum_s u^s(p)\Bar{u}^s(p) \left(\theta(p^0)f_{+}^s(\boldsymbol{p}) + \theta(-p^0)f_{-}^s(\boldsymbol{p}) \right)\,,
    \label{eq:Dfermion}
\end{equation}
\begin{equation}
    D_{\frac{1}{2}}(p)|_{\rm unpol.} = (\slashed{p} + m) \left(\frac{i}{p^2 - m^2 + i\epsilon} - (2\pi) \delta(p^2 - m^2) \left(\theta(p^0)f_{+}(\boldsymbol{p}) + \theta(-p^0)f_{-}(\boldsymbol{p}) \right) \right).
        \label{eq:Dfermion_unpolarized}
\end{equation}
The latter matches the result from \cite{Ghosh:2022nzo}. 

For the vector particles the polarized and unpolarized propagators at finite density in the $R_\xi$ gauge are found to be
\begin{align}
    D_{1}^{\mu \nu}(p)|_{\rm pol.}  &= \frac{-iP_\xi^{\mu\nu}}{p^2 + m^2 + i\epsilon} + 2\pi \delta(p^2 - m^2)\sum_i \epsilon^{\mu}_{i}(\boldsymbol{p})\epsilon^{\nu *}_{i}(\boldsymbol{p})\Big(\theta(p^0)f_{+}^i(\boldsymbol{p}) + \theta(-p^0)f_{-}^i(\boldsymbol{p}) \Big) \label{eq:Dvector}\\
    D_{1}^{\mu \nu}(p)|_{\rm unpol.}  &= - \frac{i P_\xi^{\mu\nu}}{p^2 + m^2 + i\epsilon} + 2\pi \delta(p^2 - m^2)\left(\theta(p^0)f_{+}(\boldsymbol{p}) + \theta(-p^0)f_{-}(\boldsymbol{p})  \right)  P^{\mu\nu} \,,\label{eq:Dvector_unpolarized}
\end{align}
with $i=(1,2)$ for the massless vector and $i=(1,2,3)$ for the massive vector. 
The Lorentz structure of the vacuum term is the usual one,
\be
P_\xi^{\mu \nu} = g^{\mu \nu} -(1-\xi)\frac{p^\mu p^\nu}{p^2-\xi m^2}\,,
\ee
while the Lorentz structure of the unpolarized background term is
\begin{equation}
    P^{\mu \nu} = \left\{\begin{matrix}
g^{\mu \nu} \: \: \: \: \: \: \:  \: \: \: \: \: \: \:   \mathrm{if} \: \: m = 0\\ 
g^{\mu \nu} - \frac{p^\mu p^\nu}{m^2} \: \: \mathrm{if} \: \: m \neq 0.
\end{matrix}\right.
\end{equation}
that appears via standard polarization sums.

\section{Effective Field Theory and  Potentials}
\label{se:EFT_Potentials}

We build the local effective operators describing the low-energy interactions between nucleons and particles of the hidden sector, that we refer to as dark particles. While this is a straightforward task for spin $0$ and $\frac{1}{2}$, more care is required for spin $1$.

The dark particles we consider may be self-conjugate (real scalar and vector, Majorana fermion) or not  (complex scalar and vector, Dirac fermion). In the latter case, the fields are charged under a $U(1)$ symmetry of the hidden sector, denoted  $U(1)_h$. This $U(1)_h$ symmetry may be gauged or not, this is  irrelevant for our purposes. The important feature here is rather that the global part of the $U(1)_h$ implies a $Z_2$ symmetry that constrains the interactions of the dark particles to be bilinear. 

Viewing the $U(1)_h$ as a spontaneously broken gauge symmetry is nevertheless useful to build consistent operators for the spin $1$ charged dark particle. Integrating out the heavy $U(1)_h$ gauge boson automatically generates consistent low-energy operators. We use this approach in section~\ref{se:spin1}.\,\footnote{
In this specific realization, the massive  ($Z'$-like) gauge boson of the broken $U(1)_h$ group has to be heavy enough to evade detection, such that the leading way to search for the light hidden sector is via the low-energy effective operators. Such a  limit  can always be taken. We emphasize that this realization is mostly used here to ensure self-consistent effective operators. In the following we remain agnostic to the UV completion of the EFT. 
   }

\subsection{Effective Operators}
\label{se:EFT}

   In the limit of unpolarized non-relativistic nucleons, only the interactions involving the nucleon bilinears $\bar N N,\bar N \gamma^\mu  N$ are relevant.   For simplicity, we assume a universal coupling to protons and neutrons---all our results are trivially generalized for non-universal couplings to nucleons. 
   The dark  scalar, fermion and vector fields   ---either self-conjugate or not --- are respectively denoted as $\phi$, $\chi$ and $X$.
  Results will be presented for a representative subset of interactions for  particles of each spin. We consider the following set of operators,
\small
\begin{align}  \label{eq:Leff}
{\cal O}^0_a&=\frac{1}{\,\Lambda} \bar N N |\phi|^2  \,,\quad\quad\quad\quad\quad\quad\quad {\cal O}^{\nicefrac{1}{2}}_a=\frac{1}{\,\Lambda^2} \bar N N \bar \chi \chi \,, \nonumber
\\ \nonumber
{\cal O}^0_b&=  \frac{1}{\,\Lambda^2}\, \bar N \gamma^\mu N i \phi^* \overleftrightarrow\partial_\mu  \phi
  \,,\quad \quad\quad\, {\cal O}^{\nicefrac{1}{2}}_b=\frac{1}{\,\Lambda^2} \bar N \gamma^\mu N \bar \chi \gamma^\mu \chi \,,
\\ \nonumber
{\cal O}^0_c&=\frac{1}{\,\Lambda^3} \bar N N \partial^\mu \phi^* \partial_\mu\phi\,,\quad\quad\quad\quad~
{\cal O}^{\nicefrac{1}{2}}_c=\frac{1}{\,\Lambda^2} \bar N \gamma^\mu  N \bar \chi \gamma^\mu \gamma^5\chi \,,
     \\  \nonumber
{\cal O}^{1}_a&=\frac{m^2}{\Lambda^3} \bar N N |X^\mu+\partial^\mu \pi|^2  \,, \\ \nonumber
{\cal O}^{1}_b&=\frac{1}{\Lambda^2} \, \bar N \gamma^\mu N \,\left(2\,{\rm Im}[X^*_{\mu\nu }X^\nu+ \partial^\nu(X_\mu X^*_\nu)]  
  -   \, {\rm Im}\left[ \pi \partial_\mu \pi^*\right]  - i  \partial_\mu \bar c_\pm c_\pm 
\right)\,, \\
{\cal O}^{1}_c&=\frac{1}{\Lambda^3} \bar N N |X^{\mu\nu}|^2 \,, ~\quad\quad\quad\quad\quad {\cal O}^{1}_d=\frac{1}{\Lambda^3} \bar N N X^{\mu\nu}\tilde X^{\mu\nu}\,.
\end{align}
\normalsize
where $\overleftrightarrow\partial=\overrightarrow\partial-\overleftarrow\partial$. The $\pi$ and $c, \bar c$ fields are respectively the Goldstone and ghosts accompanying $X$. 

  The ${\cal O}^s_b$ operators involve $U(1)_h$ currents, which  vanish if the dark particle is self-conjugate. The $2 {\rm Im}(X^*_{\mu\nu }X^\nu)+\ldots$ bilinear corresponds to the  $U(1)_h$ gauge current of the vector $X$. 
  The ${\cal O}^s_b$ operators can be thought has arising from integrating out the massive $U(1)_h$ gauge boson.

  We  introduce a discrete variable for whether or not the dark particle is self-conjugate:
\begin{align}
  \eta &= 
  \begin{cases}
  0 & \text{if self-conjugate}\\
  1 & \text{otherwise}
  \end{cases}
  \ .
\end{align}

  \subsection{The Spin-$1$ Operators}
\label{se:spin1}

Some of the spin-$1$ operators in the set \eqref{eq:Leff} require further explanation. 

\subsubsection{The ${\cal O}^1_a$ Operator}

This operator exists if the $X_\mu$ field is massive. If $X_\mu$  was massless, it could appear only via the field strength $X_{\mu\nu}$ due to $U(1)_h$ gauge invariance. Conversely, the massive $X_\mu$ field must be accompanied by its Goldstone boson. 
The gauge invariant combination is $\Theta^\mu\equiv X^\mu+\partial^\mu \pi$, which is thus what appears in ${\cal O}^1_a$.  The $|X^\mu+\partial^\mu \pi|^2$ bilinear can  arise from a gauged nonlinearly realized $U(1)$, for example in the decoupling limit of a spontaneously broken $U(1)_h$, with Abelian Higgs kinetic term $|D H|^2$ where $H \approx v e^{i\pi}$. 

The 2-pt correlators of $\Theta^\mu$ are gauge invariant, this fact is useful when computing diagrams.   In the general $R_\xi$ gauge we have
\be \langle\Theta(p)\Theta(-p)\rangle= 
\frac{i}{p^2-m^2}\left(-g_{\mu\nu}+\frac{p_\mu p_\nu}{m^2}\right)\, \label{eq:Theta2pt}
\ee
for any $\xi$. In the unitary gauge $\xi\to\infty$, the $\pi$ decouples and $\Theta_\mu\to X_\mu$, in which case  \eqref{eq:Theta2pt} simply becomes  the massive gauge boson propagator.

\subsubsection{The ${\cal O}^1_b$ Operator}

The ${\cal O}^{1}_b$ operator involves the $U(1)_h$ current of the vector field. 
A vector current can always be built by implementation of minimal coupling in the vector kinetic term, 
\be
|X_{\mu\nu}|\to | D_\mu X_\nu - D_\nu X_\mu |^2
\ee
with  $D_\mu X_\nu=(\partial_\mu +i g Z_\mu) X_\nu$ the $U(1)_h$ covariant derivative with gauge field $Z$. The resulting current is 
$J_\mu = ig\left( X^*_{\mu\nu}X^\nu - X_{\mu\nu}X^{\nu*} \right)$. The corresponding effective operator is 
\be
{\cal O}^1_{b,J}  =\frac{1}{\Lambda^2} \, \bar N \gamma^\mu N \,ig\left( X^*_{\mu\nu}X^\nu Z^\mu - X_{\mu\nu}X^{\nu*} Z^\mu\right)\,.
\ee

Independently from the gauge current, a vector particle also has in general a magnetic dipole moment. Hence  our dark particle may feature a $U(1)_h$ magnetic moment, described by the operator $Z_{\mu\nu}X^\mu X^{\nu*}$, or $  2i Z_\mu  {\rm Im}\left[
\partial_\nu  (X^\mu X^{\nu *})
\right]$ upon integration by part. This magnetic dipole in turn contributes to the EFT as  
\be
{\cal O}^1_{b,{\rm dipole }}  =\frac{1}{\Lambda^2} \, \bar N \gamma^\mu N \,2\, {\rm Im}\left[
\partial_\nu  (X^\mu X^{\nu *})
\right]\,.
\ee

A  method to consistently obtain both  the current and the magnetic dipole operator is to  identify the charged vector from fields living in the $SU(2)/U(1)$ coset
 of a gauged $SU(2)$ group broken to $U(1)$.\,\footnote{This is analogous to electroweak theory upon decoupling the $U(1)_B$ field via $\theta_w\to 0$.  }
   Additionally, the remaining $U(1)_h$ can be used to build the full nucleon-dark particle operator ${\cal O}^{1}_b$ upon breaking the $U(1)_h$ group and integrating out the massive gauge boson,  
    assuming that the nucleons have $U(1)_h$ charge.

The $SU(2)$ gauge Lagrangian takes the form
\begin{eqnarray}
{\cal L} &=& -\frac{1}{2}|X_{\mu\nu}|^2-\frac{1}{4}Z^2_{\mu\nu} -ig \left(
Z_{\mu\nu}X^\mu X^{\nu *} + X^*_{\mu\nu}X^\nu Z^\mu - X_{\mu\nu}X^{\nu*} Z^\mu
\right) +\ldots
\label{eq:L_SU2a}
\\
&=& -\frac{1}{2}|X_{\mu\nu}|^2-\frac{1}{4}Z^2_{\mu\nu} +2g\, {\rm Im}\left[
 X^*_{\mu\nu}X^\nu +\partial_\nu(X_\mu X_\nu^*)
\right]Z_\mu +\ldots
\label{eq:L_SU2b}
\end{eqnarray}
where the ellipses represent  quartic interactions. 
The second line simply is a  rewriting convenient in order to integrate out the $Z_\mu$ field. 
The corresponding vertex computed from these cubic terms for $Z^\mu(q)$, $W^\alpha(p_1)$, $W^\beta(p_2)$ legs is $ i g\left( g_{\mu\beta}(q-p_2)_\alpha+g_{\mu\alpha}(p_1-q)_\beta  +g_{\alpha\beta}(p_2-p_1)_\mu \right)$, as expected from a non-Abelian gauge theory.

From \eqref{eq:L_SU2a} we see that the $SU(2)/U(1)$ realization enforces a specific combination of the operators: ${\cal O}^1_{b,J} +g {\cal O}^1_{b,{\rm dipole}} $.\,\footnote{{Since  ${\cal O}^1_{b,J}$ and $ {\cal O}^1_{b,{\rm dipole}}$  are separately gauge-invariant, their linear combination is not enforced by gauge invariance. This makes clear that 
another UV realization  could in principle lead to a different combination of these two operators.} } This is the combination we choose to use in our representative set \eqref{eq:Leff}.

For consistency we have to include  the ghosts and the Goldstone bosons of the broken $SU(2)$. This  is necessary to perform the loop computation made in section~\ref{se:V1b}. The relevant interactions are
\be
{\cal L} = - ig Z^\mu \partial_\mu \bar c_\pm c_\pm  -  g Z^\mu \, {\rm Im}\left[ \pi \partial_\mu \pi^*  \right] + \ldots 
\ee
where we have included two pairs of ghosts, $c_+$ and $c_-$, 
which are  associated with the $X_\mu$ and $X_\mu^*$ fields. See e.g. \cite{Romao:2012pq} for a consistent set of Feynman rules of spontaneously broken $SU(2)$ gauge theory.

\subsection{The $V^1_b$ Quantum Potential}
\label{se:V1b}

We revisit the computation of the quantum (i.e. Casimir-Polder type) force induced by a bubble of charged spin-1 particles interacting with the nucleon via the ${\cal O}_b^1$ operator. Non-relativistic potentials are computed from QFT  by taking limits of $S$-matrix elements $i{\cal M}$ (see e.g. \cite{Feinberg:1989ps} for a review and \cite{Costantino:2019ixl}, App. B for EFT-related subtleties). 

For a massive spin-1 particle, loops of both ghosts and Goldstone bosons  contribute, 
\be
  i\mathcal{M}_b^{1,{\rm massive}} =  i\mathcal{M}_b^{X}+ i\mathcal{M}_b^{\pi} + i\mathcal{M}_b^{\rm gh}= 
 \vcenter{
        \hbox{
        \includegraphics[trim={2cm 7.7cm 2cm 6cm},clip,width=0.6\textwidth]{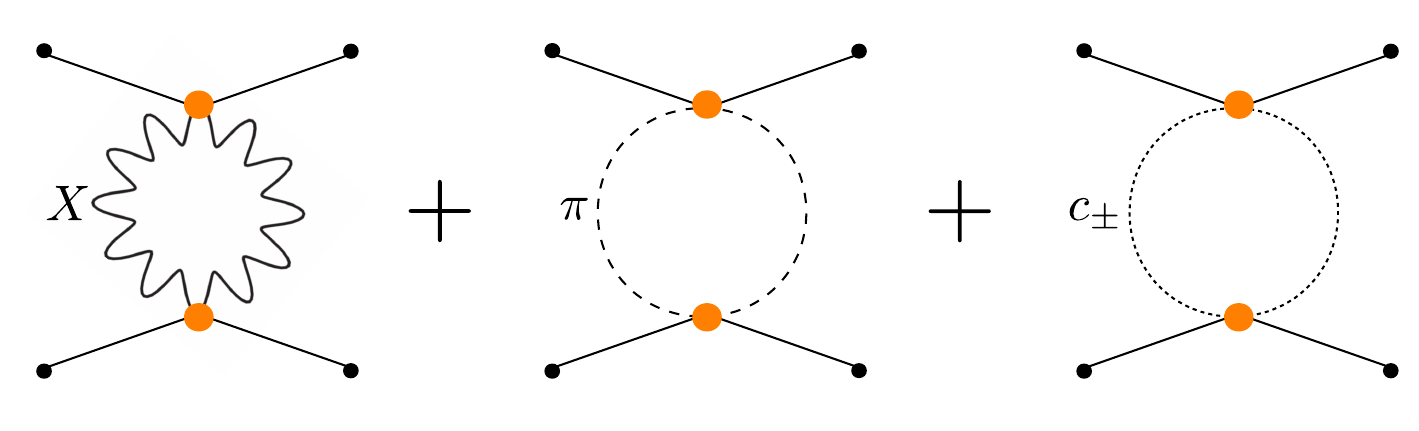}
        }
        } 
  \, \label{eq: amplitude_b_1}
\ee
The expressions for the various contributions are given in App.\,\ref{app:V1b}. 
We notice that $i\mathcal{M}_b^{\pi}=-\frac{1}{2}i\mathcal{M}_b^{\rm gh}$. This is consistent  with
the heat kernel literature, where it is known that ghosts and Goldstone contribute respectively as $-2$ and $+1$  in heat kernel coefficients \cite{Hoover:2005uf,Fichet:2013ola}. The  case of a massless spin-1 particle  is $i\mathcal{M}_b^{1,{\rm massive}} =   i\mathcal{M}_b^{X}+ i\mathcal{M}_b^{\rm gh}$.

Our calculation of $V_b^1$ corrects the one from  \cite{Fichet:2017bng}, in which the overall sign is incorrect and the Goldstone loop is not taken into account in the massive case.    We present  the results in the massless and massive case:
\be
   V_b^{1,{\rm massless}}(r) = -
   \frac{5 \eta }{8 \pi^3 \Lambda^4 r^5}\,, \label{eq:V1b_massless}
\ee
\be
    V_b^{1,{\rm massive}}(r) =  - \frac{\eta}{32 \pi^3\Lambda^4}\Bigg(\frac{36 m^3}{r^2} K_1(2 m r) + \frac{39 m^2}{r^3}K_2(2mr) \Bigg)\,, \label{eq:V1b_massive}
\ee
with  small distance limit
\begin{equation}
      V_b^{1,{\rm massive}}(r)|_{r\ll m^{-1}} = - \frac{39 \eta }{64 \pi^3 \Lambda^4 r^5}\,.
\end{equation}

The $V_b^{1,{\rm massive}}$ potential differs from $V_b^{1,{\rm massless}}$ with $m\to 0$, as expected since a theory of a massive gauge bosons does not interpolate to the massless one in the $m\to 0$ limit. 
The effect of the Goldstone loop is however only a small modification to the prediction.

These quantum potentials are attractive. Further investigation shows that  this is a consequence  of the dipole contribution. In contrast, the pure current contribution from the ${\cal O}^1_{b,J}$ operator generates a \textit{repulsive} potential 
\be
V_{b,J}^1(r)|_{r\ll m^{-1}} = \frac{3\eta}{16\pi^3\Lambda^4 r^5}\,.
\label{eq:V1bJ}
\ee
Here we show the massless result with no ghost. 
The fact that this potential is repulsive is  consistent with the argument given in \cite{brax2018bounding} that, in the vacuum,  the  potential between any    vectorial currents  is repulsive. It would be interesting to investigate how  exactly the argument is evaded in the presence of the  dipole.

\subsection{Computing the Background-Induced Potentials}
\label{se:Computations}

The leading contribution to the macroscopic force induced by the dark particle comes from a bubble diagram. The diagram is here computed at finite density using the propagators computed in section~\ref{se:propagators}. The vacuum and background terms pictured in \eqref{eq:prop_finite_density} produce the following diagrams:
\begin{align}
 \raisebox{25pt}{$i{\cal M}~=$} \vcenter{
        \hbox{
        \includegraphics[trim={0cm 3cm 0cm 6cm},clip,width=0.8\textwidth]{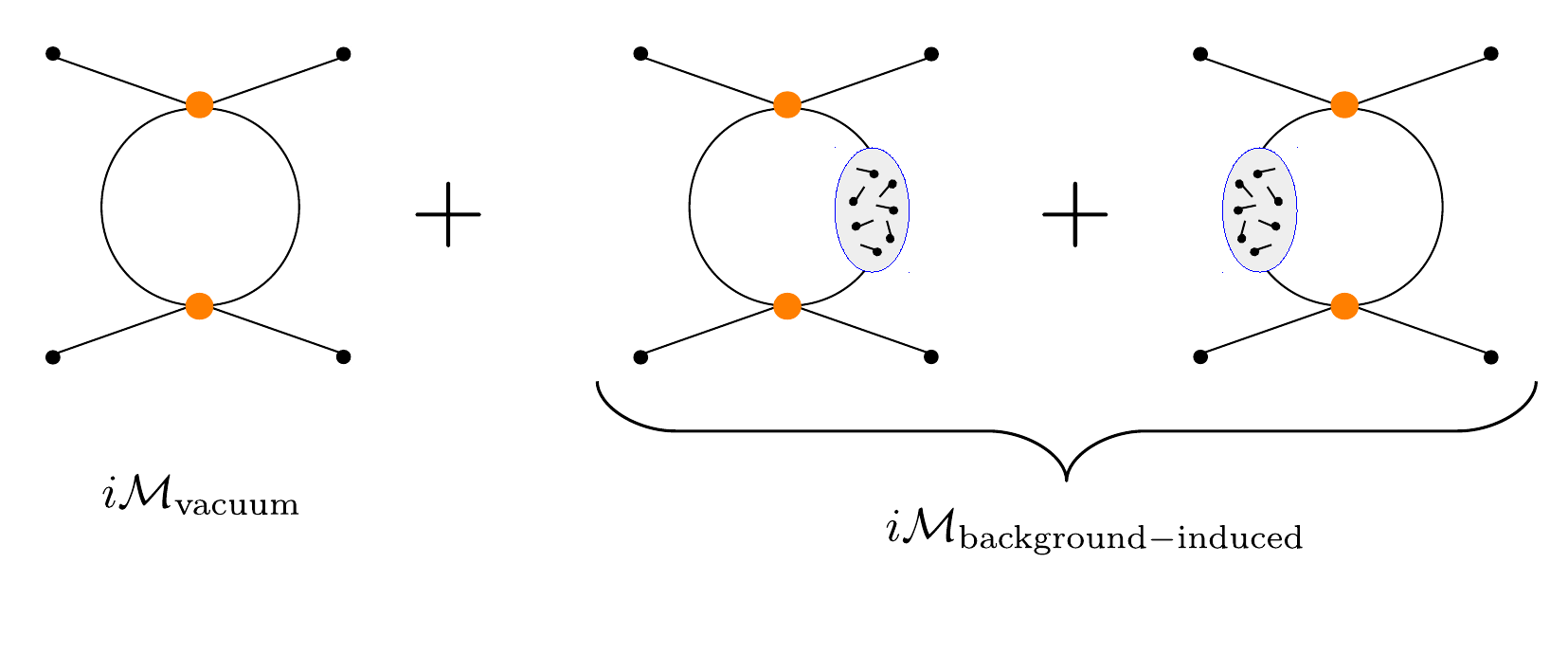}
        }
        } 
  \, \label{eq:loop_finite_density}
\end{align}
 The first term is the usual bubble producing a Casimir-Polder-like quantum potential such as the one computed in section~\ref{se:V1b}. 
The cross term is the background-induced amplitude from which we  compute the background-induced potential. 
There is also a  term featuring the squared densities, which  vanishes identically due to the presence of two delta functions and is thus not shown in \eqref{eq:loop_finite_density}. 

While the vacuum contribution is purely quantum, one may wonder about a classical view of the background-induced contribution. A classical interpretation should exist since the $i{\cal M}_{\rm background-induced}$ terms involve a single virtual line, similar to  a tree-level process. The classical understanding in terms of the sea of real particles has been recently developed in \cite{VanTilburg:2024tst},  in which context the background-induced force is appropriately called ``wake force''.

The background-induced amplitudes induced by the effective operators listed in \eqref{eq:Leff} are collected in App.\,\ref{app:amplitudes}. 
A detailed calculation for the operator $\mathcal{O}_{a}^{0}$ is given in 
App.~\ref{app:example}.  
We briefly summarize the key steps of the calculation of the potential starting from $i{\cal M}_{\rm background-induced}$.

We use the conventions from \cite{peskin2018introduction}.
We take the non-relativistic limit, for which 
$\Bar{u}^{s^\prime}(p^\prime)u^s(p) \approx 2 m \delta^{ss'} $,
$\Bar{u}^{s^\prime}(p^\prime)\gamma^\mu u^s(p) \approx 2 m \delta^{ss'} \delta^{\mu}_0$ and $q \approx (0, \bm{q})$. The definition of the non-relativistic potential is $i \mathcal{M} = -i \Tilde{V}(\boldsymbol{q}) 2 m \delta^{ss'} 2 m \delta^{r r'}$. Taking the Fourier transform leads to the spatial potential
\begin{align}
    V(\boldsymbol{r}) &= \int \frac{d^3 q}{(2\pi)^3}e^{i \boldsymbol{q}\cdot\boldsymbol{r}} \tilde{V}(\boldsymbol{q}).
\label{eq:Vdef}
\end{align}

The amplitude contains the loop integral $\int d^4 k$, we  integrate in $k^0$. 
The spatial potential takes the form
\begin{equation}
    V(\boldsymbol{r}) \propto - \int \frac{d^3 k}{(2\pi)^3} \int \frac{d^3 q}{(2\pi)^3}e^{i \boldsymbol{q}\cdot\boldsymbol{r}}\frac{(f_+(\bm{k}) + f_-(\bm{k}))}{E_k}\left[\frac{\mathcal{A}(q, k)}{\boldsymbol{q}^2 + 2 \boldsymbol{k}\cdot\boldsymbol{q} - i\epsilon} + \frac{\mathcal{A}(-q, k)}{\boldsymbol{q}^2 - 2 \boldsymbol{k}\cdot\boldsymbol{q} - i\epsilon} \right], \nonumber
\end{equation}
where the numerator $\mathcal{A}(q, k)$ depends on the interaction and the spin of the exchanged particle. Factoring the fractions above and using spherical coordinates, the $\int d^3 q$ integral can be evaluated as follows: 
\begin{align}
    & \frac{1}{(2\pi)^3}\int_0^{2\pi} d\phi \int_{-1}^{1}  d(\cos{\theta}) \int_0^\infty dq q^2 e^{i q r \cos{\theta}}\frac{\mathcal{F}(q^2, k)}{(q + 2 k \cos{\gamma} - i\epsilon)(q - 2 k \cos{\gamma} - i\epsilon)} \nonumber \\
    &= \frac{4\pi}{(2\pi)^3 r} \int_0^\infty dq \frac{q \sin{(q r)} \mathcal{F}(q^2, k)}{(q + 2 k \cos{\gamma} - i\epsilon)(q - 2 k \cos{\gamma} - i\epsilon)} \nonumber \\
    &= \frac{4\pi}{(2\pi)^3 r}\frac{1}{4 i} \oint dq \frac{q e^{i q r} \mathcal{F}(q^2, k)}{(q + 2 k \cos{\gamma} - i\epsilon)(q - 2 k \cos{\gamma} - i\epsilon)} \nonumber \\
    &= \frac{1}{4 \pi r}\cos{(2 k r \cos{\gamma})}\mathcal{F}(4 k^2 \cos^2{\gamma}, k)\,,
\label{eq: integral_Fourier}
\end{align}
where $\gamma$ is the angle between $\boldsymbol{k}$ and $\boldsymbol{q}$ and 
\begin{equation}
    \mathcal{F}(q^2, k) = \frac{1}{q}\Big[q(\mathcal{A}(q, k) + \mathcal{A}(-q, k)) - 2 k \cos{\gamma}(\mathcal{A}(q, k) - \mathcal{A}(-q, k))\Big].
\end{equation}
From the second to the third line of \eqref{eq: integral_Fourier} we analytically continue the integration to the complex plane and close the contour upward. 
The poles are $q_{\pm} = \pm 2 k \cos{\gamma} + i\epsilon$, both have a positive imaginary part hence both residues contribute to the integral. 

 It follows that the most general background-induced potential is
\begin{align}
    V(\boldsymbol{r}) \propto -\frac{1}{r} \int d^3 k \cos{(2 k r \cos{\gamma})}\frac{(f_+(\bm{k}) + f_-(\bm{k}))}{\sqrt{k^2 + m^2}}\mathcal{F}(4 k^2 \cos^2{\gamma}, k) \,.\label{eq:V_gen}
\end{align}
The angular integrals can  be done when the distributions are isotropic, $f_\pm(\bm k)=f_\pm(k)$.

\section{The Short Distance Limit  }
\label{se:ShortDistance}

We study the background-induced potentials at short distance. We assume isotropic distributions from now on, $f({\bm k})=f(k)$.

Starting from the general form \eqref{eq:V_gen}, we write the  $\cos(2kr \cos\gamma)$ as a series. A necessary condition for the integral of the series to be well defined is that the individual integral of each term be finite. The ${\cal F}$ numerator being polynomial in $k$, term-by-term finiteness is ensured  if  the phase space distribution function decays  faster than any polynomial at large $k$. 
In the rest of the section we assume that this property of the distribution function is true. 
The decay at large momentum of the phase space distribution function is controlled by a dimensionful scale denoted   $\mu$. We define the short distance limit with respect to this physical scale, $r\ll \frac{1}{\mu}$.

\begin{table}[t]
\centering
\begin{tabular}{|c|c|c|}
\cline{2-3}
 \multicolumn{1}{c|}{}  &  ~~~~$  r\ll \mu^{-1}$, $m \ll \mu$~~~~& ~~~~$r\ll \mu^{-1}$, $ m \gg \mu$~~~~  \\ \hline \hline
$V_a^0$ & $\displaystyle - \frac{n \langle k^{-1} \rangle}{4 \pi \Lambda^2 r}$ & $\displaystyle - \frac{n}{4 \pi \Lambda^2 m r}$    \\ [4pt] \hline
$V_b^0$ & $\displaystyle - \frac{\eta n \langle k \rangle}{\pi \Lambda^4 r}$ & $\displaystyle - \frac{\eta n m}{\pi \Lambda^4 r}$    \\ [4pt] \hline
$V_c^0$ & $\displaystyle - \frac{2 n \langle k^{3} \rangle }{5 \pi \Lambda^6 r}$ & $\displaystyle - \frac{n m^3}{2 \pi \Lambda^6 r}$    \\ 
[4pt]  \hline \hline
$V_a^{\frac{1}{2}}$ & $\displaystyle - \frac{n \langle k \rangle }{3 \pi \Lambda^4 r}$ & $\displaystyle - \frac{n m}{\pi \Lambda^4 r}$    \\ [4pt] \hline
$V_b^{\frac{1}{2}}$ & $\displaystyle - \frac{2 \eta n \langle k \rangle }{3 \pi \Lambda^4 r}$ & $\displaystyle - \frac{\eta n m}{\pi \Lambda^4 r}$    \\ [4pt] \hline
$V_c^{\frac{1}{2}}$ & $\displaystyle - \frac{2 n \langle k \rangle }{3 \pi \Lambda^4 r}$ & $\displaystyle - \frac{2 n \langle k^2 \rangle }{3 \pi \Lambda^4 m r}$    \\ [4pt] \hline \hline
$V_a^{1}$ & $\displaystyle - \frac{n \langle k^3 \rangle }{15 \pi \Lambda^6 r}$ & $\displaystyle  -\frac{n m^3}{4 \pi \Lambda^6 r}$    \\ [4pt] \hline
$V_b^{1}$ & $\displaystyle - \frac{7 \eta n \langle k \rangle}{18 \pi \Lambda^4 r}$ & $ - \displaystyle \frac{\eta n m}{\pi \Lambda^4 r}$    \\ [4pt] \hline
$V_c^{1}$ & $\displaystyle - \frac{32 n \langle k^3 \rangle }{15 \pi \Lambda^6 r}$ & $\displaystyle  -\frac{4 n m^3}{\pi \Lambda^6 r}$    \\ [4pt] \hline
$V_d^{1}$ & $\displaystyle - \frac{32 n \langle k^3 \rangle }{15 \pi \Lambda^6 r}$ & $\displaystyle  -\frac{32 n m \langle k^2 \rangle }{9 \pi \Lambda^6 r}$    \\ [4pt] \hline
\end{tabular}
\caption{Potentials induced by an isotropic background of particles of spin $s$ coupled to nucleons via the  ${\cal O}_i^s$ interactions, approximated in the small distance limit $r\ll \mu^{-1}$.  $\mu$ is a physical cutoff for the momentum in the phase space distribution function. 
  Real scalar, Majorana fermion, real vector have $\eta=0$. Complex scalar, Dirac fermion, complex vector have  $\eta=1$. 
\label{tab:V_smallr}
}
\end{table}

Let us study the short distance behavior of the potential. 
The domain of integration in $k$ in  \eqref{eq:V_gen} for which 
the integrand is unsuppressed is roughly $[0,\mu]$. 
In the short distance limit we have $\mu\ll r^{-1}$, therefore $k\ll r^{-1}$ in the region where the integrand is unsuppressed.  We can thus safely expand the integrand in $kr$ and truncate the expansion. The leading term of this expansion simply amounts to setting the cosine in \eqref{eq:V_gen}  to one. As a result the integrand is $r$-independent, leaving only the overall $1/r$ factor. 
This  proves the following property: 
\be
{\rm \it All~\operatorname{\it background-induced}~potentials~behave~as}~\,V(r)\propto \frac{1}{r}~\,{\rm \it at~small}~r.
\label{eq:Property}
\ee

The overall coefficient of the short distance potential contains the factor
\be
\int d^3k \frac{k^\h}{E}f(k)\,
\ee
with $\h$ an integer depending on the specific potential. 
Upon expanding the energy in the denominator for  either small or large $m$, 
this factor can be expressed as a \textit{moment} of the phase space distribution function. That is, the background-induced potentials with arbitrary phase-space distribution at small $r$ 
are expressed in terms of one specific characteristic of the phase space distribution function. Which characteristic appears depends on the value of $\h$.

We define the moments as follows. We introduce the standard number density 
\begin{equation}
    n = g \int \frac{d^3 k}{(2\pi)^3} f(k)\,
\label{eq:number}
\end{equation}
where $g=2^\eta$ for scalar, $g=2^{\eta+1}$ for fermion and $g = 3\cdot 2^\eta$ for massive vector. Using $n$  as a normalization factor, we define the $\h$-th moment of the phase space distribution function as 
\begin{equation}
    \langle k^\h \rangle = \frac{g}{n} \int \frac{d^3 k}{(2 \pi)^3} k^\h f(k) = \frac{g}{2 \pi^2 n}\int_0^\infty dk k^{\h+2} f(k)\,.
\label{eq: moment}
\end{equation}
We have  $\langle k^0 \rangle =1$ and   $\langle k \rangle $  is the mean momentum. By dimensional analysis we have $\langle k^\h\rangle~\sim\mu^\h$ since  it is the only dimensionful scale in $f$.

The fact that the effect of the phase space distribution is encapsulated into a single number is very useful for practical purposes. For any given phase space distribution, it is enough to compute the first moments to readily know all the potentials at small $r$.

We compute the potentials induced by particles of spin $0,\frac{1}{2},1$ coupled to  nucleons via the effective operators listed in \eqref{eq:Leff}.
The results are presented in Table \ref{tab:V_smallr}. 
It turns out that \textit{all} potentials  from the table describe  an attractive force. This is nontrivial because the sign of the finite density 
term in the propagators differs   for bosons and fermions, see section~\ref{se:propagators}. 
Furthermore we expect  that for quantum potentials the sign depend on the interaction with nucleons \cite{brax2018bounding}. The contribution of the dipole can also affect the sign as noticed in section~\ref{se:V1b}. 
 Here, in contrast, the results in Table \ref{tab:V_smallr}  suggest a universally attractive behavior. Such a behavior is reminiscent of the emergent
 long range forces observed  in Bose-Einstein condensates \cite{Ferrer:2000hm,Berezhiani:2018oxf} and of the Kohn-Luttinger effect in case of fermions \cite{KL_effect}. We leave the proof of this effect as an interesting open question. 

\section{Background-Induced Forces from Light Dark Relics  }
\label{se:BackgroundForcesRelics}

In this section and the following ones, we specify the phase space distribution function and study its consequences. Our focus in this section is on hypothetical cosmological relics with a \textit{homogeneous isotropic} spatial distribution through the cosmos. 

\subsection{On Thermal Equilibrium vs Non-equilibrium}

Phase space distribution functions at thermal equilibrium feature 
 Boltzmann suppression $f_{\rm thermal }(p)\sim e^{-\frac{E}{T}}$ in the high energy limit $E\gg T$, where $E=\sqrt{p^2+m^2}$. This in turn implies an exponential suppression 
$V(r)\sim e^{-2mr}$ 
in the background-induced potential for $rm \gg 1 $, $rT \gg 1$ (see e.g. \cite{Ferrer:1999ad} for a computation with neutrinos). 
This suppression is  the same as the one occurring in any quantum  potential 
beyond twice the Compton wavelength of the particles in the loop \cite{brax2018bounding}, $r\sim \frac{1}{2m}$. In short,
\be
V_{\rm background} ,~V_{\rm quantum} \propto e^{-2mr}~~~ {\rm when}~~~  r  \gg \frac{1}{m},\frac{1}{T}\,.
\ee
 The total force is thus  irremediably 
 exponentially suppressed  at long distance, at thermal equilibrium. 
Conversely, exponentially-\textit{un}suppressed background-induced potentials that satisfy $V_{\rm background}\gg V_{\rm quantum}$ in certain regimes
 may arise in systems \textit{beyond}   thermal equilibrium. 

\subsection{Dark Relic Phase Space Distribution}
 
 Departure from  thermal equilibrium  happens  all the time along the standard history of the Universe. 
 Whenever a species kinetically decouples from the thermal plasma it becomes a \textit{thermal relic} and its phase space distribution function gets frozen in the form it had at decoupling time. The phase space distribution keeps evolving but only due to the expansion of the spacetime background. 
The distributions of relics that  were non-relativistic at decoupling time feature a Boltzmann  factor $e^{-\frac{E}{T}}$. In contrast, the distribution of species that were relativistic at decoupling time 
behave as if the species was massless, with a $e^{-\frac{p}{T}}$ Boltzmann factor. 
This phenomenon happens for example for active neutrinos. 
Such relics  do become non-relativistic at later time, but this does not affect their frozen-out phase space distribution.

Assuming a homogeneous isotropic distribution, a fairly general ansatz that encapsulates the phase space distribution of these warm relics is \cite{Colombi:1995ze}
  \be
  f_{\operatorname{relic}}(k) =  \frac{a}{e^{b k /T}\pm 1}\,
  \label{eq:n_relic}
  \ee
  with $a,b$ real numbers. 
For example the distribution for each polarization of each active neutrino has $a=1$, $b=(4/11)^{1/3}$. 
The distribution \eqref{eq:n_relic} can also describe  warm dark matter, see e.g. \cite{Colombi:1995ze, Giudice:2000dp,Abazajian:2001nj}.

\subsection{On Decoupling and EFT}

\label{se:decoupling}

A condition for the phase space distribution \eqref{eq:n_relic} to be valid is that  the species be rela\-tivistic at decoupling time. 
This condition translates as a condition on the strength of the effective interaction, controlled by the $\Lambda$ parameter, as a function of the particle mass $m$. 
The decoupling temperature is estimated by equating Hubble scale $H\sim T^2/M_{\rm Pl}$ with interaction rate $\Gamma$. 

The interaction rate between the species and the thermal bath  in the EFT is 
\be\Gamma \sim \frac{T^{2n+1}}{\Lambda^{2n}} \ee
with $n=1,2,3$ depending on the effective operator considered in \eqref{eq:Leff}. 
If the EFT description remains true up to the decoupling temperature $T_{\rm dec}$, 
 the condition $\Gamma\sim H$ determines the decoupling temperature to be 
\be
T^{\rm EFT}_{\rm dec} \sim \left(\frac{\Lambda^{2n}}{M_{\rm Pl}}\right)^{\frac{1}{2n-1}}\,.\label{eq:condLambdaEFT}
\ee
Hence the  species is relativistic at decoupling if \be m< \left(\frac{\Lambda^{2n}}{M_{\rm Pl}}\right)^{\frac{1}{2n-1}} \label{eq:condmEFT}\,. \ee 
As an aside, the $T^{\rm EFT}_{\rm dec}$ scale is lower than the validity scale of the EFT, which is $\sim 4\pi\Lambda$.  Hence condition \eqref{eq:condmEFT} is stronger than the condition for EFT validity.

If the species mass is higher than $T^{\rm EFT}_{\rm dec} $, it is non-relativistic at decoupling hence the  distribution \eqref{eq:n_relic} is not well-motivated, at least in a standard cosmological scenario.  It is however possible that the EFT be UV completed at a scale \textit{below}   $T^{\rm EFT}_{\rm dec} $. The UV completion necessarily softens  the $T$-dependence. As a result the interaction rate grows \textit{slower} with $T$ in the UV completion than in the EFT. This implies in turn that  the actual $T_{\rm dec}$ can be much higher than $T^{\rm EFT}_{\rm dec}$, allowing for  higher values of $m$. 

 We conclude that it is well motivated to allow values of the mass over the full range allowed by EFT,  $m\in [0, 4\pi\Lambda]$. The only caveat being  that for masses beyond the bound \eqref{eq:condmEFT}, some (weakly-coupled) UV completion must appear at an intermediate scale. This UV completion may have other observable consequences, however here our focus is on the bounds from background-induced potentials, we do not study  correlated constraints that appear in specific UV models.

\subsection{Results}
\label{se:V_relics}

We compute the potentials induced by dark relics with the Bose-Einstein-like (BE) and Fermi-Dirac-like (FD) phase space  distributions shown in \eqref{eq:n_relic}. We report our results assuming $a=b=1$. The generalization to generic $a,b$ is straightforward. 
 We also perform the computations for a Maxwell-Boltzmann (MB) distribution in order to discern the effect of spin statistics  at low momenta $k<T$. The distributions used in the following are
\be
f_{\rm BE}(k)=\frac{1}{e^{k/T} - 1} \,,\quad \quad 
f_{\rm FD}(k) = \frac{1}{e^{k/T} + 1} \,,\quad \quad
f_{\rm MB}(k) = e^{-\frac{k}{T}}\,. \label{eq:n_dist}
\ee
The distributions are isotropic. 

We assume that the spin $1/2$ background is unpolarized and that the spin $1$ is transverse unpolarized with no longitudinal component. 
Hence we use the fermion propagator \eqref{eq:Dfermion_unpolarized}
with  $f^s(k)\equiv f_{\rm FD}(k)$ for any $s$ and  the vector propagator \eqref{eq:Dvector_unpolarized} with 
$f^i(k)\equiv f_{\rm BE}(k) $ for any $i$. 
Finally we assume that the distributions are equal for particles and antiparticles, i.e. vanishing chemical potential.

The computation of the potentials follows the steps presented in section~\ref{se:Computations}. The exact potentials are presented in integral form in App.\,\ref{app:potentials}. All potentials have the same structure with a $1/E_k$ in the integrand, with the exception of $V_b^0$ for which a cancellation
with the numerator occurs. For this reason the $V_b^0$ has a somewhat peculiar behavior in certain limits.  We compute the integrals in the $m\ll T$ and $m\gg T$ limits. The former can also be seen as  $m=0$. We further expand for $r \gg T^{-1}$ and $r \ll T^{-1}$. The resulting four limits are presented in Table \ref{tab:V_relics} for MB, BE and FD distributions.

From Tab.\,\ref{tab:V_relics} we see that for $r \ll T^{-1}$, the potentials are $1/r$, as predicted by our general result \eqref{eq:Property}.
The powers of $T$ are readily understood via dimensional analysis. 
Moreover the coefficients depends on the moments of the phase space distribution function.

We can explicitly check compatibility of these computations with the generic ones of Tab.\,\ref{tab:V_smallr}.
The  number densities obtained from \eqref{eq:number}, \eqref{eq:n_dist} are 
\begin{align}
    n_{\mathrm{MB}} = \frac{g T^3}{\pi^2}\,,\quad\quad
    n_{\mathrm{BE}} = \frac{g \zeta(3) T^3}{\pi^2} \,, \quad \quad
    n_{\mathrm{FD}} = \frac{3 g \zeta(3) T^3}{4 \pi^2}\,.
\end{align}
The first moments are computed in  Tab.\,\ref{tab:moments}. Plugging them in the results of Tab.\,\ref{tab:V_smallr} reproduces exactly the $r\ll T^{-1}$ results of Tab.\,\ref{tab:V_relics}.

\begin{table}[t]
\centering
\begin{tabular}{| c|c|c|c|c|}
\cline{2-5}
 \multicolumn{1}{c|}{}  & $\langle k^{-1} \rangle$ & $\langle k \rangle$ & $\langle k^{2} \rangle$ & $\langle k^{3} \rangle$  \\ \hline\hline
 MB & $\frac{1}{2 T}$ & $3 T$ & $12 T^2$ & $60 T^3$     \\ [2pt] \hline
 BE & $\frac{\pi^2}{12 \zeta(3)T}$  & $\frac{\pi^4 T}{30 \zeta(3)}$ & $\frac{12\zeta(5)T^2}{\zeta(3)}$ & $\frac{4 \pi^6 T^3}{63\zeta(3)}$      \\ 
 [2pt] \hline
FD & $\frac{\pi^2}{18 \zeta(3)T}$ & $\frac{7 \pi^4 T}{180 \zeta(3)}$ & $\frac{15 \zeta(5) T^2}{\zeta(3)}$ & $\frac{31 \pi^6 T^3}{378 \zeta(3)}$      \\ [2pt] \hline
\end{tabular}
\caption{First moments of the phase space distribution functions \eqref{eq:n_dist}.
  \label{tab:moments}
}
\end{table}

\begin{table}[h!]
\centering
\begin{tabular}{| c|c|c|c|c|}
\cline{2-5}
 \multicolumn{1}{c|}{}  & $m \ll T $, $r \ll T^{-1}$ & $m \ll T$, $r \gg T^{-1}$ & $m \gg T$, $r \ll T^{-1}$ & $m \gg T$, $r \gg T^{-1}$  \\ \hline\hline
 $V^0_{a,\,{\rm MB}}$ & $-\frac{2^\eta T^2}{8\pi^3 \Lambda^2 r}$ & $-\frac{2^\eta}{32 \pi^3 \Lambda^2 r^3}$ & $-\frac{2^\eta T^3}{4\pi^3 \Lambda^2 m r}$ & $-\frac{2^\eta }{64\pi^3 \Lambda^2 m T r^5}$     \\ [2pt] \hline
 $V^0_{a,\,{\rm BE}}$ & $-\frac{2^\eta T^2}{48\pi \Lambda^2 r}$  & $-\frac{2^\eta T}{32 \pi^2 \Lambda^2 r^2}$ & $-\frac{2^\eta \zeta(3)T^3}{4\pi^3\Lambda^2 m r}$ & $-\frac{2^\eta T}{32 \pi^3 \Lambda^2 m r^3}$      \\ 
 [2pt] \hline
$V^0_{b,\,{\rm MB}}$ & $-\frac{6 \eta T^4}{\pi^3 \Lambda^4 r}$ & $\frac{\eta}{8 \pi^3 \Lambda^4 r^5}$ & $-\frac{2\eta m T^3}{\pi^3 \Lambda^4 r}$ & $-\frac{\eta m}{8\pi^3 \Lambda^4 T r^5}$     \\ [2pt] \hline
$V^0_{b,\,{\rm BE}}$  & $-\frac{\eta \pi T^4}{15 \Lambda^4 r}$  & $-\frac{\eta}{16 \pi^3 \Lambda^4 r^5}$ & $-\frac{2\zeta(3)\eta m T^3}{\pi^3\Lambda^4 r}$ & $-\frac{\eta m T}{4 \pi^3 \Lambda^4 r^3}$      \\ [2pt] \hline
$V^0_{c,\,{\rm MB}}$  & $-\frac{3\cdot 2^{\eta+3} T^6}{\pi^3 \Lambda^6 r}$ & $-\frac{3 \cdot 2^{\eta-4}}{ \pi^3 \Lambda^6 r^7}$ & $-\frac{2^{\eta-1} T^3 m^3}{\pi^3 \Lambda^6 r}$ & $-\frac{2^{\eta-5} m^3}{\pi^3 \Lambda^6 T r^5}$     \\ [2pt] \hline
$V^0_{c,\,{\rm BE}}$   & $-\frac{2^{\eta+3} \pi^3 T^6}{315 \Lambda^6 r}$  & $-\frac{3\cdot 2^{\eta-3}T}{\pi^2 \Lambda^6 r^6}$ & $-\frac{2^{\eta-1}\zeta(3) m^3 T^3}{\pi^3\Lambda^6 r}$ & $-\frac{2^{\eta-4} m^3 T}{\pi^3 \Lambda^6 r^3}$      \\ [2pt] \hline \hline 
$V^{\nicefrac{1}{2}}_{a,\,{\rm MB}}$  & $-\frac{2^{\eta-1} 4 T^4}{\pi^3 \Lambda^4 r}$ & $\frac{ 2^{\eta-1} 3}{4 \pi^3 \Lambda^4 r^5}$ & $-\frac{ 2^{\eta-1} 4 m T^3}{\pi^3 \Lambda^4 r}$ & $-\frac{2^{\eta-1} m}{4 \pi^3 \Lambda^4 T r^5}$     \\ [2pt] \hline
$V^{\nicefrac{1}{2}}_{a,\,{\rm FD}}$   & $-\frac{2^{\eta-1} 7 \pi T^4}{180 \Lambda^4 r}$  & $\frac{2^{\eta-1} 3}{8 \pi^3 \Lambda^4 r^5}$ & $-\frac{2^{\eta-1} 3 \zeta(3) m T^3}{\pi^3 \Lambda^4 r}$ & $-\frac{2^{\eta-1} m}{16 \pi^3 \Lambda^4 T r^5}$      \\ [2pt] \hline
$V^{\nicefrac{1}{2}}_{b,\,{\rm MB}}$ & $-\frac{8\eta T^4}{\pi^3 \Lambda^4 r}$ & $-\frac{\eta}{2 \pi^3 \Lambda^4 r^5}$ & $-\frac{4\eta T^3 m}{\pi^3 \Lambda^4 r}$ & $-\frac{\eta m}{4 \pi^3 \Lambda^4 T r^5}$     \\ [2pt] \hline
$V^{\nicefrac{1}{2}}_{b,\,{\rm FD}}$  & $-\frac{7\eta \pi T^4}{90 \Lambda^4 r}$  & $-\frac{\eta}{4 \pi^3 \Lambda^4 r^5}$ & $-\frac{3\eta \zeta(3) m T^3}{\pi^3\Lambda^4 r}$ & $-\frac{\eta m}{16 \pi^3 \Lambda^4 T r^5}$      \\ [2pt] \hline
$V^{\nicefrac{1}{2}}_{c,\,{\rm MB}}$  & $-\frac{2^{\eta+2} T^4}{\pi^3 \Lambda^4 r}$ & $-\frac{2^{\eta-2}}{\pi^3 \Lambda^4 r^5}$ & $-\frac{2^{\eta+4} T^5}{\pi^3 \Lambda^4 m r}$ & $-\frac{2^{\eta-2}}{\pi^3 \Lambda^4 m T r^7}$     \\ [2pt] \hline
$V^{\nicefrac{1}{2}}_{c,\,{\rm FD}}$   & $-\frac{7\cdot 2^{\eta-2} \pi T^4}{45 \Lambda^4 r}$  & $-\frac{2^{\eta-3}}{\pi^3 \Lambda^4 r^5}$ & $-\frac{15 \zeta(5) \cdot 2^{\eta} T^5}{\pi^3 \Lambda^4 m r}$ & $-\frac{2^{\eta-4}}{\pi^3 \Lambda^4 m T r^7}$      \\ [2pt] \hline\hline
$V^{1}_{a,\,{\rm MB}}$ & $-\frac{3 \cdot 2^{\eta+2} T^6}{\pi^3 \Lambda^6 r}$ & $-\frac{15 \cdot 2^{\eta-4}}{ \pi^3 \Lambda^6 r^7}$ & $-\frac{3\cdot 2^{\eta-2} T^3 m^3}{\pi^3 \Lambda^6 r}$ & $-\frac{3\cdot 2^{\eta-6} m^3}{\pi^3 \Lambda^6 T r^5}$     \\ [2pt] \hline
$V^{1}_{a,\,{\rm BE}}$  & $- 
    \frac{2^{\eta+2} \pi^3 T^6}{315 \Lambda^6 r}$  & $-\frac{3 \cdot 2^{\eta-4} T}{\pi^2 \Lambda^6 r^6}$ & $-\frac{3 \cdot 2^{\eta-2} \zeta(3) m^3 T^3}{ \pi^3 \Lambda^6 r}$ & $-\frac{3 \cdot 2^{\eta-5} m^3 T}{ \pi^3 \Lambda^6 r^3}$      \\ [2pt] \hline
$V^{1}_{b,\,{\rm MB}}$  & $-\frac{7 \eta T^4}{\pi^3 \Lambda^4 r}$ & $-\frac{19\eta}{16 \pi^3 \Lambda^4 r^5}$ & $-\frac{6\eta T^3 m}{\pi^3 \Lambda^4 r}$ & $-\frac{3 \eta m}{8 \pi^3 \Lambda^4 T r^5}$     \\ [2pt] \hline
$V^{1}_{b,\,{\rm BE}}$   & $-\frac{7 \eta \pi T^4}{90 \Lambda^4 r}$  & $-\frac{\eta T}{2 \pi^2 \Lambda^4 r^4}$ & $-\frac{6\eta\zeta(3)T^3 m}{\pi^3\Lambda^4 r}$ &$-\frac{3\eta m T}{4 \pi^3 \Lambda^4 r^3}$      \\ [2pt] \hline
$V^{1}_{c,\,{\rm MB}}$  & $-\frac{3 \cdot 2^{\eta+7} T^6}{\pi^3 \Lambda^6 r}$ & $-\frac{15\cdot 2^{\eta+1}}{ \pi^3 \Lambda^6 r^7}$ & $-\frac{3 \cdot 2^{\eta+2} T^3 m^3}{\pi^3 \Lambda^6 r}$ & $-\frac{3 \cdot 2^{\eta-2} m^3}{\pi^3 \Lambda^6 T r^5}$     \\ [2pt] \hline
$V^{1}_{c,\,{\rm BE}}$   & $-\frac{2^{\eta+7} \pi^3 T^6}{315 \Lambda^6 r}$  & $-\frac{3\cdot 2^{\eta+1} T}{ \pi^2 \Lambda^6 r^6}$ & $-\frac{3 \cdot 2^{\eta+2}\zeta(3)T^3 m^3}{\pi^3\Lambda^6 r}$ &$-\frac{3 \cdot 2^{\eta-1} m^3 T}{\pi^3 \Lambda^6 r^3}$      \\ [2pt] \hline
$V^{1}_{d,\,{\rm MB}}$  & $-\frac{3 \cdot 2^{\eta+7} T^6}{\pi^3 \Lambda^6 r}$ & $-\frac{15\cdot 2^{\eta+1}}{ \pi^3 \Lambda^6 r^7}$ & $- \frac{2^{\eta+7} m T^5}{\pi^3 \Lambda^6 r}$ & $\frac{5 \cdot 2^{\eta+1} m}{\pi^3 \Lambda^6 T r^7}$     \\ [2pt] \hline
$V^{1}_{d,\,{\rm BE}}$   & $-\frac{2^{\eta+7} \pi^3 T^6}{315 \Lambda^6 r}$  & $-\frac{3\cdot 2^{\eta+1} T}{ \pi^2 \Lambda^6 r^6}$ & $-\frac{2^{\eta+7}\zeta(5)T^5 m}{\pi^3\Lambda^6 r}$ &$\frac{3 \cdot 2^{\eta+1} m T}{\pi^3 \Lambda^6 r^5}$      \\ [2pt] \hline
\end{tabular}
\caption{The potentials $V^s_i$ induced from the ${\cal O}^s_i$ effective operators, assuming  Maxwell-Boltzmann (MB), Bose-Einstein (BE) or Fermi-Dirac (FD) distribution for the relics.
  \label{tab:V_relics}
}
\end{table}

The key feature of the $m\gg T$, $r \gg T^{-1}$ regime is that the force is \textit{not} exponentially suppressed, unlike its quantum counterpart. This has important consequences for experimental searches, as discussed in section~\ref{se:5thForces}. 

The behavior at $m\ll T$, $r \gg T^{-1}$ also presents a pattern. For the FD and MB distributions the temperature vanishes. Due to this feature the $r$-dependence can be deduced from dimensional analysis. For the BE distribution, in all cases except the $V_b^0$ potential which features a cancellation, the result is \textit{enhanced} by a factor $ T r$ with respect to the MB case. This is a manifestation of the bosonic statistics, that becomes relevant at long distances  i.e. low momenta $k \ll T$. 
For example, the $V_0^a$ potential computed with the BE distribution behaves as $T/r^2$ while it behaves as $1/r^3$ with the MB distribution.\,\footnote{
Fifth forces behaving as $1/r^3$ and $1/r^2$ are also predicted from AdS and linear dilaton braneworld models, see respectively \cite{ Callin:2004py} and \cite{Fichet:2022xol}.
Forces with $1/r^2$ behavior can also be generated from mixing to a CFT sector 
\cite{Brax:2019koq,Costantino:2019ixl, Chaffey:2021tmj}. }

Potentials induced by Dirac and Majorana neutrinos have been computed in \cite{Ghosh:2022nzo}. These potentials are simply related to our $V_b^{\frac{1}{2}}$, $V_c^{\frac{1}{2}}$ potentials. Our results agree upon appropriate translation of conventions. 

{An analysis of the experimental bounds is presented in section \ref{se:5thForces}, where we assume a temperature close to that of the cosmic microwave background. }

\FloatBarrier

\section{Background-Induced Forces from Cold Dark Matter}
\label{se:BackgroundForcesDarkMatter}

Our focus in this section is on cosmological relics with a virialized distribution, analogous to  cold dark matter.

\subsection{The Standard Halo Model}

Astronomical observations suggest the existence of halos of non-baryonic dark matter  surrounding galaxies. The local velocity distribution of the dark matter particle is commonly described using an isothermal distribution known as the Standard Halo Model (SHM) \cite{Drukier:1986tm}. The SHM is confirmed by precise cosmological hydrodynamic simulations of galaxy formation such as those in \cite{Kelso:2016qqj,Sloane:2016kyi,Bozorgnia:2016ogo}.  
The SHM assumes a Maxwell-Boltzmann distribution with a velocity dispersion $v_0\ll 1$ that corresponds to the rotational velocity of the disk in the solar neighborhood, with a cut-off at velocities exceeding the galaxy escape velocity $v_{\mathrm{esc}}$,
\begin{equation}
     \tilde{f}(v) = \begin{cases}
\frac{1}{N_{\mathrm{esc}}}\left(\pi v_0^2 \right)^{-3/2} \; e^{-v^2/v_0^2}, & v < v_{\mathrm{esc}}  \\
0, & v > v_{\mathrm{esc}} 
\end{cases}
\end{equation}
with normalization 
\begin{equation}
    N_{\mathrm{esc}} = \mathrm{erf}(z) - \frac{2}{\sqrt{\pi}} z e^{-z^2}\,,\quad\quad z = \frac{v_{\mathrm{esc}}}{v_0}\,.
\label{eq:SHM}
\end{equation}

\subsection{Dark Matter Phase Space Distribution}

The dark relic of our interest may be identified as the dark matter particle itself. It could also be identified as  a subcomponent, contributing to only a fraction of the observed mass-energy content of the universe.

We assume that the dark matter density $\rho_0$ in the Solar neighborhood is constant, and use that the distribution \eqref{eq:SHM} forces the dark particle to be non-relativistic, such that $\rho_0 = m n_0$ with $n_0$ the number density and $m$ the dark relic  mass.
The phase space distribution is then deduced from the SHM \eqref{eq:SHM} as follows:
\begin{equation}
     f(k) = \begin{cases}
\frac{a}{N_{\mathrm{esc}}}\left(\frac{2 \sqrt{\pi}}{m v_0} \right)^3 \frac{n_0}{g} \; e^{-\frac{k^2}{m^2 v_0^2}}, & k < m v_{\mathrm{esc}}  \\
0, & k > m v_{\mathrm{esc}} \,.
\end{cases}\label{eq:DM_distribution}
\end{equation}
 The overall factor $a\leq 1$ takes the value $a=1$ if the dark relic is identified as dark matter, while  if $a<1$  it is identified as a subcomponent contributing with fraction $a$ to the total relic density.
The distribution is normalized such that applying the definition of the number density \eqref{eq:number} gives 
\begin{equation}
    g \int \frac{d^3 k}{(2 \pi)^3} f(k) = a\, n_0 =a\frac{\rho_0}{m} \,.
\end{equation}



\subsection{Results}

\label{se:V_DM}

We compute the potentials induced by the dark matter distribution \eqref{eq:DM_distribution}. We report our results assuming $a=1$. 
From the exact expressions presented in integral form in App.\,\ref{app:potentials}, we calculate the potentials in the short and large distance limits, $r \ll (m v_0)^{-1}$ and $r \gg (m v_0)^{-1}$, respectively. 
The results are summarized in Table \ref{tab:V_DM}.

 All the potentials are attractive. 
Notice that, due to the normalization of \eqref{eq:DM_distribution},  the results do not depend on the $2^\eta$ factors, that get automatically absorbed to produce the $n_0$ density. 

In the short distance limit, our results  are consistent with those presented in Tab.\,\ref{tab:V_smallr}, identifying $\mu \equiv m v_0$. Hence the force present a universal $1/r$ behavior. In this context, only the second column of the Table \ref{tab:V_smallr} is relevant, since $v_0 \ll 1$. 
{In the large distance limit, the potentials exhibit exponential suppression, similar to the discussion in \ref{se:BackgroundForcesRelics}, except that here the transition occurs at  a distance  $r\sim \frac{1}{m v_0}$, that is  much larger than the Compton wavelength since $v_0\ll1$. }
Other general features of the potentials have been already noted in section \ref{se:V_DM}, the discussion similarly applies here. 
A detailed example of computation using the distribution \eqref{eq:DM_distribution} is given in App.~\ref{app:example}.

The anisotropy due to the Earth's motion relative to the rest frame of the dark matter halo is  discussed in section \ref{sec: anisotropy}.


\begin{table}[t]
\centering
\begin{tabular}{|c|c|c|}
\cline{2-3}
 \multicolumn{1}{c|}{}  &  ~~~~$  r\ll (m v_0)^{-1}$, $v_0 \ll 1$~~~~& ~~~~$r\gg (m v_0)^{-1}$, $ v_0 \ll 1$~~~~  \\ \hline \hline
$V_a^0$ & $\displaystyle - \frac{n_0 }{4 \pi \Lambda^2 m r}$ & $\displaystyle - \frac{n_0 }{4 \pi \Lambda^2 m r} e^{- r^2 m^2 v_0^2}$    \\ [4pt] \hline
$V_b^0$ & $\displaystyle - \frac{\eta n_0 m}{\pi \Lambda^4 r}$ & $\displaystyle - \frac{\eta n_0 m}{\pi \Lambda^4 r} e^{- r^2 m^2 v_0^2}$    \\ [4pt] \hline
$V_c^0$ & $\displaystyle - \frac{n_0 m^{3} }{2 \pi \Lambda^6 r}$ & $\displaystyle - \frac{n_0 m^{3} }{2 \pi \Lambda^6 r} e^{-r^2 m^2 v_0^2}$    \\ 
[4pt]  \hline \hline
$V_a^{\frac{1}{2}}$ & $\displaystyle - \frac{n_0 m}{\pi \Lambda^4 r}$ & $\displaystyle -\frac{n_0 m}{\pi \Lambda^4 r} e^{-r^2 m^2 v_0^2}$    \\ [4pt] \hline
$V_b^{\frac{1}{2}}$ & $\displaystyle - \frac{\eta n_0 m }{ \pi \Lambda^4 r}$ & $\displaystyle - \frac{\eta n_0 m}{\pi \Lambda^4 r} e^{- r^2 m^2 v_0^2}$    \\ [4pt] \hline
$V_c^{\frac{1}{2}}$ & $\displaystyle - \frac{n_0 m v_0^2}{\pi \Lambda^4 r}$ & $\displaystyle - \frac{n_0 m v_0^2 }{\pi \Lambda^4 r} e^{-r^2 m^2 v_0^2}$    \\ [4pt] \hline \hline
$V_a^{1}$ & $\displaystyle - \frac{n_0 m^3 }{4 \pi \Lambda^6 r}$ & $\displaystyle  - \frac{n_0 m^3 }{4 \pi \Lambda^6 r} e^{-r^2 m^2 v_0^2}$    \\ [4pt] \hline
$V_b^{1}$ & $\displaystyle - \frac{\eta n_0 m}{\pi \Lambda^4 r}$ & $ \displaystyle - \frac{\eta n_0 m}{\pi \Lambda^4 r} e^{- r^2 m^2 v_0^2}$    \\ [4pt] \hline
$V_c^{1}$ & $\displaystyle - \frac{4 n_0 m^3}{\pi \Lambda^6 r}$ & $\displaystyle   - \frac{4 n_0 m^3}{\pi \Lambda^6 r} e^{-r^2 m^2 v_0^2}$    \\ [4pt] \hline
$V_d^{1}$ & $\displaystyle - \frac{16 n_0 m^3 v_0^2}{3 \pi \Lambda^6 r}$ & $\displaystyle  -\frac{16 n_0 m^3 v_0^2}{3 \pi \Lambda^6 r} e^{-r^2 m^2 v_0^2}$    \\ [4pt] \hline
\end{tabular}
\caption{Potentials induced from the  ${\cal O}_i^s$ interactions, assuming the dark matter distribution \eqref{eq:DM_distribution} for the  relics.
\label{tab:V_DM}
}
\end{table}

\FloatBarrier

\section{ Bounding  Dark Relics Using Fifth Force Searches}
\label{se:5thForces}

We investigate experimental bounds on the background-induced potentials computed in section~\ref{se:BackgroundForcesRelics} and summarized in Tab.\,\ref{tab:V_relics}. 
A broad landscape of existing fifth force searches has been recast in \cite{brax2018bounding}, see also \cite{Banks:2020gpu}. 
We revisit some of the bounds from \cite{brax2018bounding}, and also provide a few minor corrections.

\subsection{Experiments with Planar Geometry}
\label{se:planar}

A number of sub-millimeter scale fifth force experiments have tested the attraction between two dense objects with approximate planar  geometries \cite{Lamoreaux:1996wh,Fischbach:2001ry,Smullin:2005iv,Decca:2005qz,Decca:2007jq,Decca:2007yb,bordag2009advances, Adelberger:2009zz, Chen:2014oda}. 
Here we use the results from  \cite{Chen:2014oda} (IUPUI),  \cite{Fischbach:2001ry} (AFM), \cite{Smullin:2005iv} (Stanford) 
and also discuss the E\"ot-Wash experiment \cite{Adelberger:2003zx}. 

The basic dimensions and materials of each setups are collected in \cite{brax2018bounding}, Tab.\,2.

\subsubsection{Computing the Observables}
\label{se:planarforces}

\paragraph{Proximity Force Approximation}

When two objects are close enough, the   proximity force (or Derjaguin's) approximation (PFA) \cite{bordag2009advances} applies: The force is proportional to the potential energy between the two objects approximated as infinite planes. Computing the potential between two layered planes is possible using the piecewise mass density function describing $n$ layers over a bulk with density $\rho$,
\be\gamma(z)=
\begin{cases}
\rho_n \quad &{\rm if}\quad 0<z<\Delta_n \\
\rho_{n-1} \quad &{\rm if}\quad \Delta_n<z<\Delta_n+\Delta_{n-1}
\\ \,\,\vdots
\\ \rho \quad &{\rm if}\quad z>\sum_i^n\Delta_i \,
\end{cases}\,
\ee
where  layer $n$ is the closest to the other plate \cite{brax2018bounding}.
The potential between an infinite plate of (possibly piecewise) density $\gamma_a(z)$ and a plate with area $A$ and density $\gamma_b(z)$ at separation $s$ is 
\be
V_i^{\rm plate}=2\pi A \int_0^\infty d\rho\, \rho \int_0^\infty dz_a \gamma_a(z_a) \int_0^\infty  dz_b \gamma_b(z_b) \, V_i(\sqrt{\rho^2+(s+z_a+z_b)^2})\,. \label{eq:Vplane}
\ee

\paragraph{IUPUI and AFM.} In these two experiments the geometry is sphere-plane with small separation. The PFA applies to a very good accuracy. We compute the forces using \eqref{eq:Vplane}. In the IUPUI  experiment the observable is the difference of force between a test mass and two kinds of materials --- gold and silicon.

\paragraph{Stanford.} In the Stanford experiment the separation between the sample and the plane is not small enough to apply the PFA. We rather approximate the geometry as point-plane. The observable is a difference of forces between the test mass and planar areas of gold and silicon.

\subsubsection{A Consistency Check} 

The geometries used above are mere approximations of the actual experimental setup. The accuracy of the approximation   deserves to be verified. 
We propose the following method. 

We refer to the observable as $O$. It can be  a force, a difference  of forces, a torque, etc depending on the experiment.
Our theoretical prediction is $O_{\rm th}$.  
The experiments traditionally release a bound on a Yukawa-like force  with potential $V_{\rm Yuk} \propto \alpha \frac{e^{-mr}}{r}$ in the form of an exclusion region in the $(m,\alpha)$ plane. We denote the boundary of the exclusion region as $\alpha_{\rm ex}(m)$.

In order to check whether our predicted observable $O_{\rm th}$ actually reflects the quantity probed by the experiment, we compute it in the Yukawa case, $O^{\rm Yuk}_{\rm th}(s,m)$ and inspect the product
\be
\alpha_{\rm ex}(m) O^{\rm Yuk}_{\rm th}(s_{0},m) \,
\ee
where $s_{0}$ is the nominal separation of the experiment. This quantity provides a nontrivial check of the approximations used to compute $O_{\rm th}$.

 If our modelization of the geometry is indeed correct, then the $\alpha_{\rm ex}(m) O^{\rm Yuk}_{\rm th}(s_{0},m)$ pro\-duct must be approximately constant in $m$, up to some uncertainty inherent to the  rough recasting we are doing. 
We can then safely use the nearly constant value of $\alpha_{\rm ex}(m) O^{\rm Yuk}_{\rm th}(s_0,m)$ to bound the observable $O_{\rm th}(s)$  for any other non-Yukawa force and for any separation $s$. 
Conversely, if $\alpha_{\rm ex}(m) O^{\rm Yuk}_{\rm th}(s_{0},m)$ is not constant in $m$, we cannot trust our approximation of the geometry,  and therefore we cannot use the bound.

We apply this validation method to the IUPUI \cite{Chen:2014oda}, AFM \cite{Fischbach:2001ry,bordag2009advances} and Stanford \cite{Smullin:2005iv}  bounds on Yukawa forces. We  find good agreement, the results are summarized in Tab.~\ref{tab:planar}.

\begin{table}[t]
    \centering
    \begin{tabular}{|c|c|c|c|c|}
    
       \cline{2-5}
 \multicolumn{1}{c|}{}  &  IUPUI \cite{Chen:2014oda} & AFM \cite{Fischbach:2001ry,bordag2009advances} & Stanford 1 \cite{Smullin:2005iv} & Stanford 2  \cite{Smullin:2005iv}  \\
         \hline 
      $\log_{10}\left(\frac{\alpha_{\rm ex}(m) O^{\rm Yuk}_{\rm th}(s_0,m)}{{\rm eV}}\right)$   & $15.8+O(0.1)$ & $21.3+O(0.1)$ & $15.5+O(1)$ & $15.0+O(1)$  \\
      \hline
    \end{tabular}
    \caption{
   The  nearly-constant value of the product $\alpha_{\rm ex}(m) O^{\rm Yuk}_{\rm th}(s_0,m)$ for any $m$ within the tested mass range  for each of the  experiments considered, using the geometries given in section~\ref{se:planarforces}.
     We use these values to constrain the    $O_{\rm th}(s,m)$ observables. 
    }
    \label{tab:planar}
\end{table}

\subsubsection*{The E\"ot-Wash experiment}

Applying the same consistency check to the bound on Yukawa-like force from the   E\"ot-Wash experiment
does not seem to provide conclusive results. Our $\alpha_{\rm ex}(m)O^{\rm Yuk}_{\rm th}(s_0,m)$ function varies over many orders of magnitude if one uses  the approximate torque given in \cite{Adelberger:2009zz} as $O^{\rm Yuk}_{\rm th}$.  
Taking a conservative approach leads to a bound similar to the Stanford 1 experiment. 
Hence we do not include the E\"ot-Wash bound in this work. 
However, given the tendencies observed from the other experimental bounds, the E\"ot-Wash result may provide the leading bound, hence it would be good to properly recast it.\,\footnote{
In  \cite{brax2018bounding} we used the E\"ot-Wash results only for the $1/r^3$ potential, for which it plays an important role. To this end we directly used the bound provided in \cite{Adelberger:2006dh}.   }

\subsection{Molecular Bounds}
\label{se:molecules}

Impressive progress on both the experimental
\cite{Niu201444,Biesheuvel:2016azr,UbachsAPB, Balin2011, Hori2011,PhysRevLett.98.173002,PhysRevLett.108.183003,Ubachs09}  and the theoretical \cite{Karr14,PhysRevLett.113.023004,Karr2016,KHK14,PhysRevA.76.022106,PhysRevA.82.032509, 0953-4075-37-11-010,PhysRevA.74.052506,PhysRevA.77.042506,PhysRevA.77.022509,doi:10.1021/ct900391p,Pachu11}
 sides of precision molecular spectroscopy  have been accomplished in the past decade,
opening the possibility of searching for extra forces below the \AA~scale using  transition frequencies of well-understood  molecular systems. Certain of these results have recently been used to bound short distance modifications of gravity, see Refs.\,\cite{PhysRevD.87.112008,Salumbides:2013dua,Ubachs17,Ubachs20161}.

The most relevant systems for which both precise measurements and predictions are available are the hydrogen molecule H$_2$, the molecular hydrogen-deuterium ion HD$^+$, the antiprotonic helium $\bar p\,^4$He$^+$ and muonic molecular deuterium ion $dd\mu^+$, where $d$ is the deuteron. 

 The presence of an extra force shifts the energy levels of these systems. 
In \cite{brax2018bounding} we computed the energy shifts for the transitions between the  $(\nu=1,J=0)-(\nu=0,J=0) $ states for H$_2$, the $(\nu=4,J=3)-(\nu=0,J=2) $ states of HD$^+$, the  $(m=33,l=32)-(m=31,l=30) $ states of $\bar p \,^4$He$^+$, and the binding energy of the $(\nu=1,J=0)$ state of $dd\mu^+$ using the wave functions given in \cite{Salumbides:2013dua,Ubachs17}.

The bounds on the extra forces can then be obtained by asking that $\Delta E$ be smaller than the combined (theoretical + experimental) uncertainties $\delta E$. 
We reproduce the uncertainties  in Tab.~\ref{tab:deltaE} (see references for details).\,\footnote{A few corrections of the values have been done with respect to \cite{brax2018bounding} and \cite{Banks:2020gpu}. 
}
\begin{table}[t]
    \centering
    \begin{tabular}{|c|c|c|c|c|c|}
    \hline
      System   & H$_2$ & HD$^+$ & $\bar p^4{\rm He}^+$ &  $dd\mu^+$ \\
      \hline
        $\delta E$ & 3.2\,neV \cite{Ubachs17,PhysRevD.87.112008}  & 0.33\,neV \cite{Ubachs17,PhysRevD.87.112008}
        & 4.0\,neV \cite{Salumbides:2013dua} & 0.7\,meV \cite{Salumbides:2013dua} \\
        \hline
    \end{tabular}
    \caption{Combined uncertainties for molecular spectroscopy measurements.}
    \label{tab:deltaE}
\end{table}

\FloatBarrier

\subsection{Bounds on Light Dark Relics}

\label{se:Bounds_light_relics}

We present the exclusion regions on the  potentials induced by light dark relics with the homogeneous backgrounds  described in section \ref{se:BackgroundForcesRelics} and Tab.\,\ref{tab:V_relics}.  We assume the temperature is $T=2.7 $ K. 
{ Under this assumption, the transition between the heavy and light mass regimes used in Tab.~\ref{tab:V_relics} is at $m\sim 10^{-4}$ eV while the transition between short and long distance regimes is at  $r \sim 1$ cm.} 


According to the standard lore of structure formation, such relics should have a mass lower than  $1-10$ keV. 
 Keeping open the possibility of loopholes or alternative mechanism to evade this upper bound, we extend the domain of the plot to an order of magnitude higher.

\begin{figure}[t]
\centering
\includegraphics[trim={0cm 0cm 0cm 0cm},clip,width=0.7\textwidth]{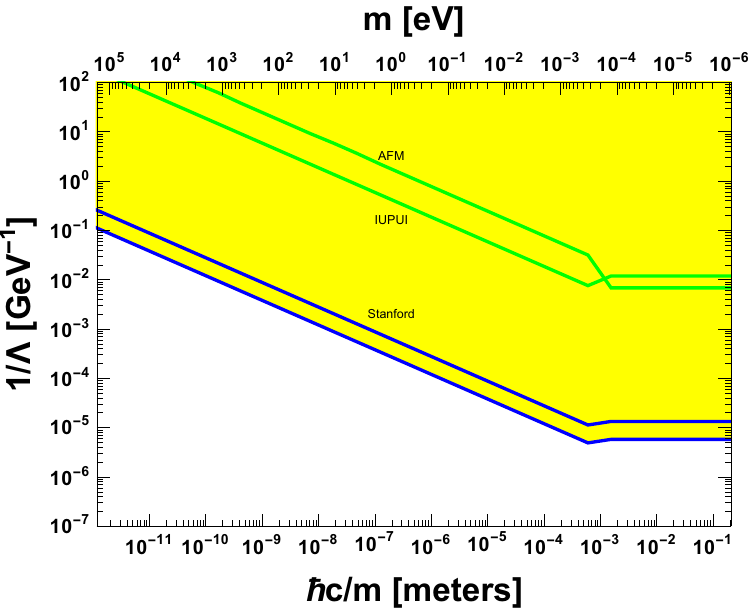} 
\caption{\it   
Bounds on the  potential $V_a^0$ induced by a real scalar relic ($\eta=0$)  
with  Bose-Einstein distribution
coupled to nucleons via the ${\cal O}_a^0$ interaction. 
 The yellow region is excluded. Details on the experimental bounds are given in the text and \cite{brax2018bounding}. 
Below the dotted line, relativistic decoupling can occur within the EFT regime, while above the dotted line relativistic decoupling  occurs in a suitable UV completion (see section~\ref{se:decoupling}). 
}
\label{fig:V0a}
\end{figure}

\begin{figure}[t]
\centering
\includegraphics[trim={0cm 0cm 0cm 0cm},clip,width=0.7\textwidth]{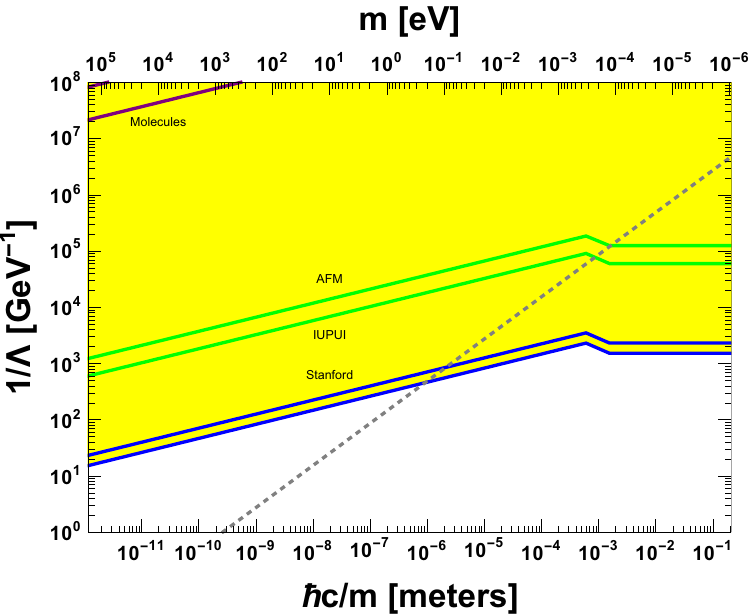} 
\caption{\it   
Bounds on the  potential $V_b^0$ induced by   a complex scalar relic ($\eta=1$) with  Bose-Einstein distribution  coupled to nucleons via the ${\cal O}_b^0$ interaction. Same conventions as in Fig.\,\ref{fig:V0a}. 
}
\label{fig:V0b}
\end{figure}

\begin{figure}[t]
\centering
\includegraphics[trim={0cm 0cm 0cm 0cm},clip,width=0.7\textwidth]{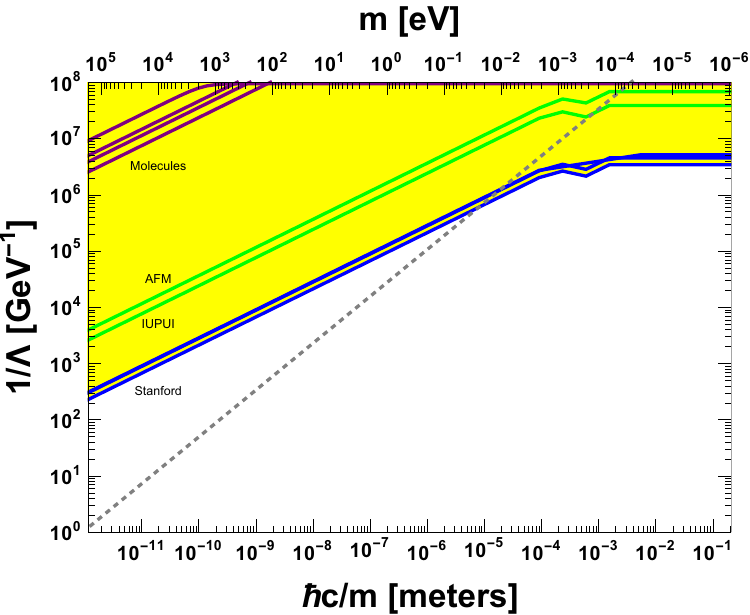} 
\caption{\it   
Bounds on the  potential $V_c^0$ induced by   a real scalar relic ($\eta=0$) with  Bose-Einstein distribution  coupled to nucleons via the ${\cal O}_c^0$ interaction. Same conventions as in Fig.\,\ref{fig:V0a}. 
}
\label{fig:V0c}
\end{figure}

\begin{figure}[t]
\centering
\includegraphics[trim={0cm 0cm 0cm 0cm},clip,width=0.7\textwidth]{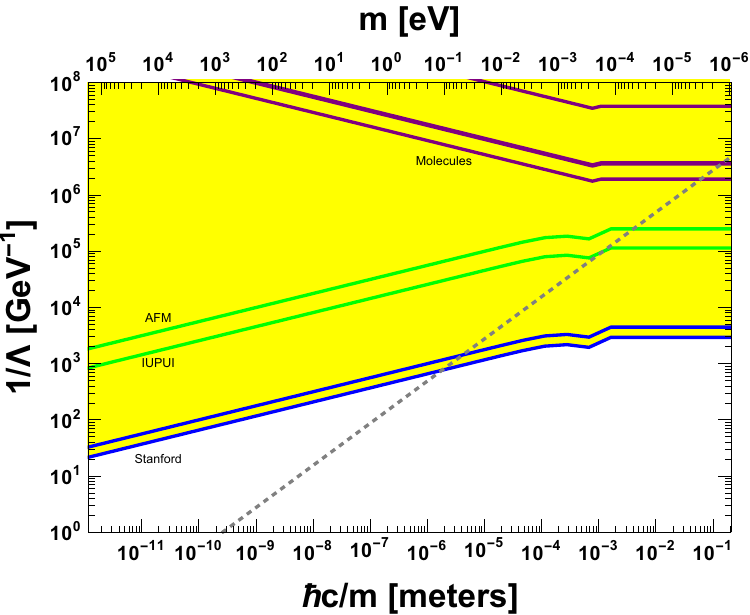} 
\caption{\it   
Bounds on the  potential $V_b^{\frac{1}{2}}$ induced by  a Dirac relic with  Fermi-Dirac distribution($\eta=1$)  coupled to nucleons via the ${\cal O}_b^{\frac{1}{2}}$ interaction. Same conventions as in Fig.\,\ref{fig:V0a}. 
}
\label{fig:V12b}
\end{figure}

We focus on the potentials $V^0_{a}$, $V^0_{b}$, $V^0_{c}$, $V^{\frac{1}{2}}_b$  in Figs.\,\ref{fig:V0a},\,\ref{fig:V0b},\,\ref{fig:V0c},\,\ref{fig:V12b} respectively. 
The bounds on the other potentials are similar to these representative cases. This is because potentials generated by the spin$-1$ particle behave similarly to those induced by the scalar, and potentials generated by the spin-$1/2$ particle  behave  similarly to $V^{\frac{1}{2}}_b$ in most regimes (see Tab.\,\ref{tab:V_relics}).

\begin{figure}[t]
\centering
\includegraphics[trim={0cm 0cm 0cm 0cm},clip,width=0.7\textwidth]{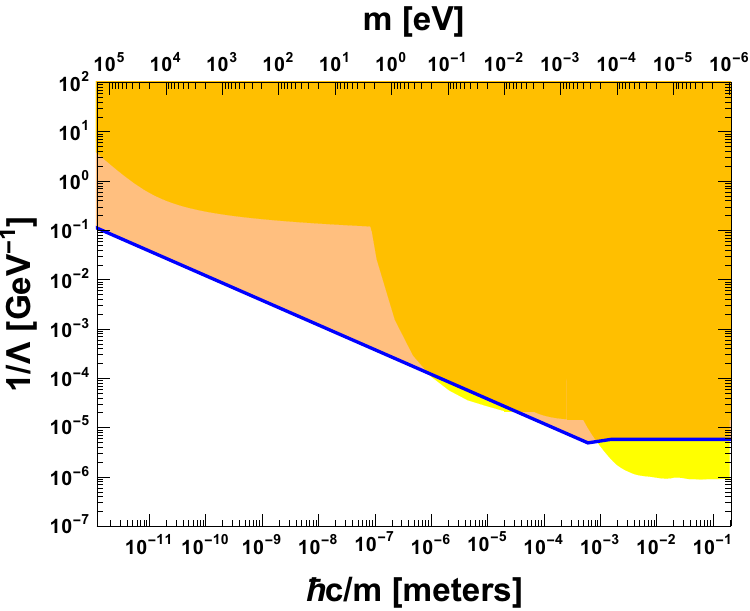} 
\caption{\it   
Comparison of the bounds on the background-induced potential $V_a^0$ shown in Fig.\,\ref{fig:V0a} to the bounds on its quantum counterpart (yellow) from \cite{brax2018bounding}. 
}
\label{fig:Comparison}
\end{figure}

A striking overall conclusion is that, in all cases, the bounds from  large-scale experiments  are stronger than those from lower scales. Here the Stanford experiment, which has the largest separation ($\sim 25~\mu$m) among the experimental setups considered, dominates for any $m$ in each case.
This is in sharp contrast with the bounds on the quantum potentials, which can be often dominated by short-scale experiments depending on $m$ and on the steepness of the potential (see \cite{brax2018bounding}).
The feature of large-scale dominance is tied to the fact that the background-induced forces from dark relics with the thermal distributions given in \eqref{eq:n_dist}
are not exponentially suppressed at long distance. 
As a result there is no hierarchy between the bounds as a function of the mass.

The shape of the exclusion landscape is overall quite different from the one on quantum forces from \cite{brax2018bounding}. Our conventions are the same as in this reference, a direct comparison  can be done.
In the $V_a^0$ plot it turns out that the background-induced potential excludes more parameter space than the quantum contribution in the mass region above $m\sim 1 $\,eV. This is shown in Fig.\,\ref{fig:Comparison}. This mass range includes warm dark matter scenarios.

In the other cases, the background-induced bound is typically of  same order of magnitude as the quantum one. Their nature is however very different: the leading bound on the quantum force comes from molecular spectroscopy and neutron scattering while the bound on the background-induced component comes from the Stanford experiment.

Given the tendency of large-scale dominance, it would be very interesting to reliably compute the E\"ot-Wash bound, which may provide the strongest of all fifth force bounds for any $m$.

We remind that in the  case of the $V_b^{\frac{1}{2}}$ potential  the contribution from the background-induced potential cancels the quantum contribution in the large distance limit~\cite{Ferrer:1999ad}. This feature  is an exception rather than  a pattern, it does not happen in the other cases. Hence for our purposes it is sufficient to  focus only on the background-induced part of the potentials.

All our exclusion regions are computed for the nominal phase space distribution functions of \eqref{eq:n_dist}. The bounds for a generic overall coefficient $a$ are easily obtained by rescaling the couplings. The rescaling factors for the four plots are respectively $\Lambda\to\Lambda a^{-1/2}$, $\Lambda\to \Lambda a^{-1/4}$, $\Lambda\to \Lambda a^{-1/6}$, $\Lambda\to \Lambda a^{-1/4}$.

{Finally we mention that light thermal dark relics are  constrained by cosmological observations, see e.g.  \cite{Dvorkin:2022jyg} for a review. Such bounds rely on the standard  model of cosmology, hence it is in principle possible that modified cosmological scenarios evade such constraints. The bounds we provide here are of a different nature, since they are reproducible experimental bounds that only depend on the local phase space distribution today.  We thus present them independently of the cosmological scenario, with no strong prior on the domain of the parameters. }

\FloatBarrier

\subsection{Bounds on Scalar Dark Matter}

\label{se:Bounds_dark_matter}

We present the exclusion regions on the  potentials induced by scalar  dark relics identified as cold dark matter, using the SHM distribution \eqref{eq:DM_distribution}. We use $\rho_0= 0.58$ GeV cm$^{-3}$, $v_0= 220$ km s$^{-1}$
\cite{Benito:2020lgu, Sloane:2016kyi}. 
We focus on the $V_a^0$ potential produced by the ${\cal O}^0_a$ interaction (see Tab.\,\ref{tab:V_DM}), for which the strongest bounds appear. 

{The common lore of structure formation indicates that cold dark matter should have mass higher  than the keV scale. 
However, loopholes may be possible, see for example \cite{Moroi:2020has, Bessho:2022yyu,Yin:2023jjj,Choi:2023jxw}, 
in which bosonic dark matter down to the eV scale is found to be cold. For this reason we present our result extended to lower masses, down to the eV scale. }

{While the bounds on $\Lambda$ for masses near the keV are weak, they grow much stronger at lower masses. 
For example, at $m=1$\,eV the bound from the Stanford experiment reaches $\Lambda\gtrsim 10^3$\,TeV. 
Converting to a coupling with the typical UV form $\frac{1}{\Lambda}\equiv\frac{m_N}{f^2}$,  we have that $f\gtrsim 1$\, TeV at $m=1$\,eV. 
At lower masses, the appropriately recasted E\"ot-Wash results would most likely produce a strong bound,  as discussed in section \ref{se:Bounds_light_relics}.  }


\begin{figure}[t]
\centering
\includegraphics[trim={0cm 0cm 0cm 0cm},clip,width=0.7\textwidth]{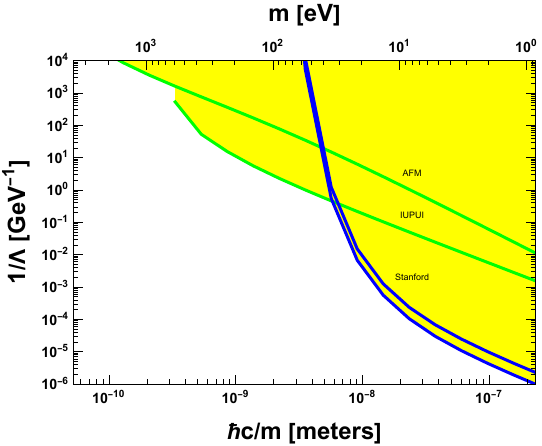} 
\caption{\it   
Bounds on the  potential $V_a^0$ induced by   a real scalar relic ($\eta=0$) identified as dark matter, coupled to nucleons via the ${\cal O}_a^0$ interaction. Same conventions as in Fig.\,\ref{fig:V0a}. 
}
\label{fig:V0a_DM}
\end{figure}


\section{Anisotropy from Earth's Motion}\label{sec: anisotropy}

Throughout this paper we have assumed an isotropic phase space distribution. Here we investigate a realistic anisotropic case. 

From the cosmological viewpoint, it is reasonable to assume  
the dark relics  to be at rest relative to the CMB.
 Measurements of CMB anisotropy reveal that the Earth is moving at approximately $370$ km/s relative to the CMB frame \cite{Planck:2018nkj}. 
 The Earth's motion can induce anisotropy such that the momentum will receive a Galilean boost, $f(\bm{k}) \rightarrow f(\bm{k^\prime}) = f(\bm{k} + m \bm{u})$, being $\bm{u}$ the Earth's velocity relative to the relic rest frame.

To evaluate the effect of the anisotropy in the general expression \eqref{eq:V_gen}, we perform a variable substitution $\bm{k}^\prime = \bm{k} + m \bm{u}$ such that
\begin{align}
    V(\boldsymbol{r}) &\propto -\frac{1}{r} \int d^3 k^\prime \cos{(2 \bm{k}^\prime \cdot \bm{r} - 2 m \bm{u}\cdot \bm{r})}\frac{(f_+(\bm{k}^\prime) + f_-(\bm{k}^\prime))}{\sqrt{(\bm{k}^\prime - m \bm{u})^2 + m^2}}\mathcal{F}(\bm{k}^\prime, \bm{u}, \hat{r})  
    \,.\label{eq:V_ani}
\end{align}

We introduce the angle $\theta$ between the Earth's motion and the separation vector of the pair of nucleons feeling the potential, 
\(\bm{u} \cdot \bm{r} = u r \cos{\theta}\) with $r=|{\bm r}|$, $u=|{\bm u}|$. We choose the frame such that ${\bm u}= u \hat z$. 

The potential has a strong dependence on the \(\theta\) angle. The integral over the angles is not straightforward to evaluate, as the integrand depends on the angles between the vectors $\bm{k}^\prime$ and $\bm{u}$, as well as between $\bm{k}^\prime$ and $\bm{r}$. However, assuming that $u \ll 1$, which is  true e.g. in the Earth's motion case,  the dependence on the angle between $\bm{k}^\prime$ and $\bm{u}$ becomes negligible. 

\begin{figure}[t]
\centering
\includegraphics[trim={0cm 0cm 0cm 0cm},clip,width=0.6\textwidth]{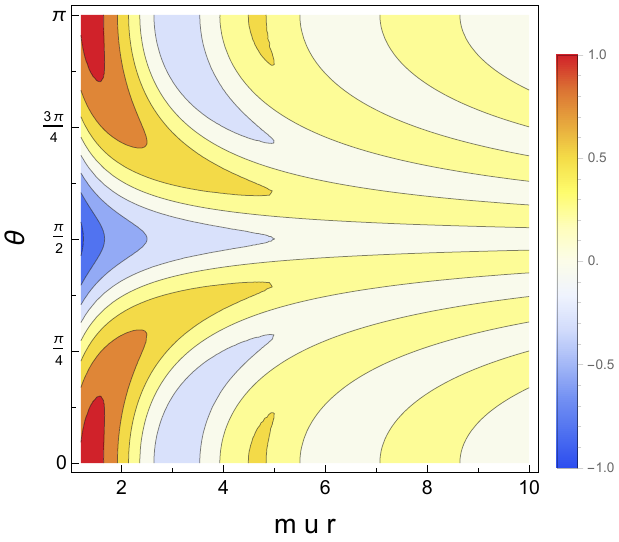} 
\caption{\it Effect of anisotropy on potentials in the short-distance limit. The potentials depend on the angle $\theta$ between  the separation vector of the two nucleons, $\bm r $, and  Earth's velocity, $\bm u$.
}
\label{fig:V_anisotropy}
\end{figure}

In the short distance limit, $r \ll T^{-1}$,  the potential takes the form
\begin{equation}
    V(\bm{r}) \propto - \frac{1}{r} \cos{(2 m u r \cos{\theta})}\,.
\end{equation}
{We present this result for any $\theta$ in Fig.\,\ref{fig:V_anisotropy}. When the nucleon pair is oriented perpendicularly to the Earth's velocity ($\theta = \pi/2$), the potential reduces to the isotropic result in the short distance limit (Tab. \ref{tab:V_smallr}.)
For arbitrary orientation, the potential features a characteristic oscillating behavior  as a function of the distance.   Such a behavior could in principle be used to distinguish the signal from the background in fifth force experiments.
Additionally,  a seasonal variation is expected in the case of  anisotropy due to Earth's motion,  that  could be used to demonstrate the nature of the anisotropy. }


\FloatBarrier

\section{Summary \label{se:conclusion}}

We investigated the forces  between macroscopic bodies that are generated in the presence of  a background of real particles quadratically coupled to matter. These background-induced forces accompany the universal Casimir-Polder-like forces that already exist at zero temperature.  
While this phenomenon is general, our focus in this work is ultimately on hypothetical dark particles that may exist in a hidden sector beyond the Standard Model.

We presented the most general propagators for scalar, fermion and vectors in the presence of a --- possibly polarized --- phase space distribution function.  
The interactions with nucleons are described via a low-energy EFT.  We carefully derived from a $SU(2)/U(1)$ coset the operators describing the spin-1 dark particles interactions including magnetic moment,  ghosts and Goldstone fields. As an aside we computed the quantum potential induced by the spin-1 particle coupled to nucleons via a current operator. The quantum force is found to be repulsive, consistent with general results. In contrast, when the magnetic dipole is included, the resulting force turns out to be attractive. These results supersede a previous calculation from \cite{Fichet:2017bng}.

Focusing on isotropic distributions, we have shown that the background-induced potentials at short distance behave always as $1/r$, irrespective of details of spin, interaction and phase space distribution. Furthermore their strength is controlled by a specific moment of the phase space distribution function. 
This is useful since for any given phase space distribution function, computing  the first moments readily provides all the small-$r$ potentials. 
It turns out that all computed potentials are attractive, a nontrivial fact given the variety of particles and interactions considered.
This universal behavior  is reminiscent of the emergent
 long range forces observed  in Bose-Einstein condensates \cite{Ferrer:2000hm,Berezhiani:2018oxf} and of the Kohn-Luttinger effect in case of fermions \cite{KL_effect}. It would be interesting to prove this effect generally, we leave this as an intriguing open question.

{Backgrounds that are at thermal equilibrium induce forces that are exponentially suppressed at distances beyond the Compton wavelength of the exchanged mediator $r \sim \frac{1}{m}$, just like their quantum counterpart. In contrast, large effects can occur at distances beyond the Compton wavelength in systems away from equilibrium. As an application we
consider cosmological relics, focusing on \textit{(i)} dark species that were relativistic at the decoupling time --- analogous to SM neutrinos --- and \textit{(ii)}  dark species with virialized distribution --- analogous to, or identified as, cold dark matter. }



We first consider dark relics that were relativistic at decoupling time (case \textit{(i)}), assuming either Bose-Einstein and Fermi-Dirac-like distribution functions.
We assume isotropic unpolarized densities with vanishing chemical potential.  Our results are consistent with previous results in the literature, such as in \cite{Ghosh:2022nzo}, which  are typically focused on the neutrino  effective Lagrangian.

We find that the potential in the $r\ll T^{-1}$ regime behaves universally as $\propto 1/r$ and is proportional to moments of the thermal distribution for both $m\gg T$ and $m\ll T$, as expected from our general analysis.
The potentials in the $m\ll T$, $r\gg T^{-1}$ limit also presents a simple pattern controlled by dimensional analysis. An enhancement by a large factor $T r $ occurs in the Bose-Einstein case, due to the enhancement of the distribution at low momentum. Finally, a key feature of the  $r\gg m^{-1}$  limit is that the forces are {\it not} exponentially suppressed, unlike their quantum counterpart. 
A striking implication is that powerful large scale experiments can be used to search for dark relics of any mass within the EFT validity range, i.e. below the GeV scale.

{We turn to virialized dark relics (case \textit{(ii)}), using a phase space distribution based on the Standard Halo Model. The normalization is chosen to be smaller or equal to the one of cold dark matter. Hence our approach applies to cases where the dark relic under consideration is either identified as the dark matter of the Universe or is a subcomponent of it.  
The potentials in the $r\ll (m v_0)^{-1}$ regime behave universally as $\propto 1/r$. 
For $r\gg (m v_0)^{-1}$, the potentials are exponentially suppressed. Note however that this occurs at a distance much larger than  the Compton wavelength $r \sim m^{-1}$. 
}

We recast, based on previous works, bounds from molecular spectroscopy, from expe\-riments in the $O(10-100)$ nanometer scale and in the $O(10)$ micrometer scale (Stanford). We show numerically that the Stanford experiments dominates for any mass, for any dark particle spin and interaction. Along the process we present a method to validate the approximation of the geometry used for each experiment. Based on this, we did not attempt to recast the E\"ot-Wash experiment.

{Applying the recast experimental results to the homogeneous densities (case \textit{(i)}), }  we obtain that the bounds on dark particles from the background-induced forces typically compete with those from quantum forces. For example, the exclusion region on $\Lambda$ from the $V_a^0$ background-induced potential is stronger than the one from the quantum one for $m>1$\,keV. An important difference lies in the nature of these leading bounds: the background-induced forces are most constrained by large-scale experiments, while which bound on the quantum forces dominates  depends crucially on the dark particle mass.

{Applying the experimental results to dark matter-like densities,  the most significant bounds we find are in the case of a scalar relic with non-derivative couplings, on which we focus on. 
While the bounds on $\Lambda$ for masses near the keV are weak, they become much more significant at lower masses.  For example, at $m=1$\,eV the bound from the Stanford experiment reaches $\Lambda\gtrsim 10^3$\,TeV. 
}

{In our presentation of results, the masses range in either benchmark phase space distribution considered  extend beyond the most common expectation from the standard cosmological scenario. 
This is because loopholes based on modifications of the cosmological scenarios are always possible, to make e.g., cold dark matter lighter than the keV scale as in \cite{Moroi:2020has, Bessho:2022yyu,Yin:2023jjj,Choi:2023jxw}. In contrast, our experimental bounds from background-induced forces also have the advantage of solely depending on the local phase space distribution at present time. Hence they provide independent constraints on cosmological scenarios that evade the standard dark matter mass ranges.  }

Given the tendencies in the bounds from background-induced forces obtained in this work, the results from the E\"ot-Wash experiment  should most likely produce a powerful additional bound. It would be very interesting to properly recast these results ---  therefore turning  E\"ot-Wash into a dark relic search experiment.

Finally, we recall  that at large enough separation, the dark field propagator tends to be repelled from the sources. This is because a quadratic coupling induces an effective mass for the field propagating inside  a source. This effect has been studied in details in \cite{Brax:2022wrt, Brax:2018grq}, and has been recently dubbed ``dark matter Meissner effect" \cite{Day:2023mkb}.
At the level of the vacuum component of the force, this effect induces the transition from Casimir-Polder to Casimir-like behavior \cite{Brax:2022wrt, Brax:2018grq}. 
While the effect can be neglected in certain fifth force observables, it  must be taken into account in others, like in the phase shifts measured in cold atom interferometry \cite{Brax:2022wrt}. 
The consequences  of this ``dark Meissner effect''  on the background-induced component of the force remain to be studied. In other words, it would be certainly interesting to extend the background-induced forces of the present work into the Casimir regime.



\paragraph{Comparison to Ref.\,\cite{VanTilburg:2024tst}. 
}  This recent paper from  Ken van Tilburg's shares a similar  theme to our present work.   
While our studies are overall complementary, some of the results overlap. 
The diagrams induced by the operators ${\cal O}_{a}^0$ and ${\cal O}_{a}^{1/2}$ have been computed there. 
Notably, a  bound on  the $V_a^0$ potential is presented in the context of a DM model. The obtained sensitivity and profile of the exclusion region  are roughly similar to those from our Fig.\,\ref{fig:V0a}.

\begin{acknowledgments}

We thank Chee Sheng Fong, Flip Tanedo and Ken Van Tilburg   for useful discussions.  
This work is supported by the Sao Paulo Research Foundation (FAPESP), project 2021/10128-0.

\end{acknowledgments}

\newpage

\appendix

\section{Deriving the  Propagators at Finite Density}\label{app:propagators}

We compute the propagator for a complex scalar, then generalize to fields with spin.

\subsection{Scalar}

We calculate the propagator for a complex scalar field $\phi$ in the background state $\ket{w}$:  
\be
D_0(x-y)=\bra{w}T \phi(x)\phi^{*}(y)\ket{w}\,.
\ee
When there are antiparticles, as is the case here, the background state is a direct product of particle and antiparticles states,  $\ket{w} = \ket{w_+}\ket{w_-}$. The normalization is $\bra{w}\ket{w} = \bra{w_+}\ket{w_+} = \bra{w_-}\ket{w_-}=1$. 

We use standard canonical quantization
\begin{align}
    \phi(x) &= \int \frac{d^3 p}{(2\pi)^3} \frac{1}{\sqrt{2 E_{\bm p}}} \big(a_{\bm p} e^{-i p \cdot x} + b_{\bm p}^{\dagger} e^{ip\cdot x} \big)\\
    \phi^{*}(x) &= \int \frac{d^3 p}{(2\pi)^3} \frac{1}{\sqrt{2 E_{\bm p}}} \big(a_{\bm p}^{\dagger} e^{i p \cdot x} + b_{\bm p} e^{-ip\cdot x} \big)
\end{align}
with commutators  $[a_{\bm p}, a_{\bm k}^{\dagger}] = (2\pi)^3 \delta^{(3)}(\boldsymbol{p} - \boldsymbol{k})$ and $[b_{\bm p}, b_{\bm k}^{\dagger}] = (2\pi)^3 \delta^{(3)}(\boldsymbol{p} - \boldsymbol{k})$ for creation and annihilation operators. All other commutators vanish. 

The propagator $\bra{w}T \phi(x)\phi^{*}(y)\ket{w}$ contains the terms  $\bra{w}a_{\bm p} a_{\bm k}^{\dagger}\ket{w} = (2\pi)^3 \delta^{(3)}(\bm{p}-\bm{k}) + \bra{w_+}a_{\bm k}^{\dagger} a_{\bm p} \ket{w_+}$ and $\bra{w}b_{\bm p}^{\dagger} b_{\bm k} \ket{w} = \bra{w_-}b_{\bm p}^{\dagger} b_{\bm k} \ket{w_-}$. While the expectation value of the normal ordered terms would vanish in the vacuum, here it takes the form
\begin{align}
    \bra{w_+}a_{\bm k}^{\dagger} a_{\bm p} \ket{w_+} &= (2\pi)^3 \delta^{(3)}(\bm{p}-\bm{k})f_+(\bm{p})\,, \\ \bra{w_-}b_{\bm k}^{\dagger} b_{\bm p} \ket{w_-} &= (2\pi)^3 \delta^{(3)}(\bm{p}-\bm{k})f_-(-\bm{p}).
\end{align}

These expressions can be understood as follows. 
The $a_{\bm p}$ destroys a particle with momentum $\bm{p}$ and the $a_{\bm k}^{\dagger}$ creates a particle with momentum $\bm{k}$.  
That is, the background absorbs then emits a real particle. 
There must be momentum conservation in this process, that is ensured by the delta function.

When ${\bm k}={\bm p}$,  $a_{\bm p}^{\dagger} a_{\bm p}$ is the operator that gives the number of particles per unit volume in momentum space. From the exponential representation of the delta function we have $(2\pi)^3 \delta^{(3)}(0) \equiv V$ where $V=\int d^3x$ is the spatial volume. The  $f(\bm{p})$ factor is thus identified   as the function that gives the number of particles per unit volume in phase space, i.e. the \textit{phase space distribution function}.

Using the above definitions the propagator takes the following form:
\begin{equation}
    D_0(x-y) = \int \frac{d^3 p}{(2\pi)^3}\frac{1}{2 E_{\bm p}} \Bigg[\Big(1 + f_+(\boldsymbol{p}) \Big)e^{-i p (x-y)} + f_-(-\boldsymbol{p})e^{i p (x-y)} \Bigg].
\end{equation}
The first term is just the vacuum propagator. The second term has the same exponential factor as the first term, and can be calculated from the same contour integral. The third term has an exponential factor with opposite sign,  we  change  $\boldsymbol{p} \rightarrow -\boldsymbol{p}$ and close the contour opposite to the vacuum term. 

We perform a 4d Fourier transform to obtain the momentum space
propagator. The final result for  the  propagator of a  complex scalar at finite density is 
\begin{equation}
    D_{0}(p) = \frac{i}{p^2 - m^2 + i\epsilon} + 2\pi \delta(p^2 - m^2)[\theta(p^0)f_{+}(\boldsymbol{p}) + \theta(-p^0)f_{-}(\boldsymbol{p})]\,.
\label{eq: propagator_scalar}
\end{equation}
For a real scalar, the same calculation gives 
\begin{equation}
    D_{0}(p) = \frac{i}{p^2 - m^2 + i\epsilon} + 2\pi \delta(p^2 - m^2)f(\boldsymbol{p})\,.
\end{equation}

\subsection{Fermion}

The propagator of the Dirac fermion is defined as
\be
D_{\frac{1}{2}}(p) =  \bra{w}T\psi(x)\Bar{\psi}(y)\ket{w}\,.
\ee
We define the background state as a direct product of states with spin $s_1$ and  $s_2$, $\ket{w} = \ket{w^{s_1}} \ket{w^{s_2}}$. Following  steps  analogous to those of  the scalar case we get
\begin{align}
    &D_{\frac{1}{2}}(p) \label{eq: propagator_2} \\ &= \int\frac{d^3 p}{(2\pi)^3}\frac{1}{2 E_{\bm p}}\sum_s\Bigg[u^s(p)\Bar{u}^s(p)\Big(1 - f_+^s(\boldsymbol{p})\Big)e^{-ip(x-y)} + v^s(p)\Bar{v}^s(p) f_-^s(-\boldsymbol{p})e^{ip(x-y)}  \Bigg]  \nonumber
\end{align}
where we use the anticommutators $\{a_{\bm p}^s, a_{\bm k}^{s^\prime \dagger}\} = (2\pi)^3 \delta^{(3)}(\boldsymbol{p} - \boldsymbol{k})\delta^{s s^\prime}$ and $\{b_{\bm p}^s, b_{\bm k}^{s^\prime \dagger}\} = (2\pi)^3 \delta^{(3)}(\boldsymbol{p} - \boldsymbol{k})\delta^{s s^\prime}$. 
Going to momentum space,  we obtain the propagator for the Dirac fermion at finite density 
\begin{equation}
    D_{\frac{1}{2}}(p) =  \frac{i (\slashed{p} + m)}{p^2 - m^2 + i\epsilon} - (2\pi) \delta(p^2 - m^2)\sum_s u^s(p)\Bar{u}^s(p) \left[\theta(p^0)f_{+}^s(\boldsymbol{p}) + \theta(-p^0)f_{-}^s(\boldsymbol{p}) \right] \,.
\end{equation}
In the last term of \eqref{eq: propagator_2}, after making the change $\boldsymbol{p} \rightarrow - \boldsymbol{p}$ and since $p^0 = -E_p$, we  use $v^s(-p)\Bar{v}^s(-p) = - u^s(p)\Bar{u}^s(p)$.

If the distributions  for each spins are equal,  we recover \eqref{eq:Dfermion_unpolarized} since
$\sum_s u^s(p)\Bar{u}^s(p) = (\slashed{p} + m)$.
This unpolarized fermion propagator matches the one derived in \cite{Ghosh:2022nzo}.

\subsection{Vector}
For a vector field we follow the same steps as presented for the fermion, summing over the physical polarizations of the vector. The results are \eqref{eq:Dvector} and \eqref{eq:Dvector_unpolarized}.

\section{Computing the $V_b^1$ Quantum Potential  }
\label{app:V1b}

The diagrams from the $X$, $c_\pm$ and $\pi$ fields are respectively 
\begin{align}
  i\mathcal{M}_b^V
  = & -\frac{\eta}{\Lambda^4}(\bar{u}_{p_1'}\gamma^\mu u_{p_1}\bar{u}_{p_2'}\gamma^\nu u_{p_2})\int \frac{d^4k}{(2\pi)^4} \nonumber \\ &
  \frac{ g_{\mu\beta}(k+2q)_\alpha+g_{\mu\alpha}(k-q)_\beta  +g_{\alpha\beta}(-2k-q)_\mu}{k^2-m^2}  \nonumber
  \\ &
\frac{g_{\nu\beta}(k+2q)_\alpha+g_{\nu\alpha}(k-q)_\beta  +g_{\alpha\beta}(-2k-q)_\nu}{(p+k)^2-m^2}
\end{align}
\be
  i\mathcal{M}_b^{gh}
  =   2 \frac{\eta}{\Lambda^4}(\bar{u}_{p_1'}\gamma^\mu u_{p_1}\bar{u}_{p_2'}\gamma^\nu u_{p_2})\int \frac{d^4k}{(2\pi)^4}
    \frac{k^\mu}{k^2-m^2}  \frac{  (k+q)^\nu}{(p+k)^2-m^2}
\ee
\be
  i\mathcal{M}_b^{G}
  = - \frac{1}{4} \frac{\eta}{\Lambda^4}(\bar{u}_{p_1'}\gamma^\mu u_{p_1}\bar{u}_{p_2'}\gamma^\nu u_{p_2})\int \frac{d^4k}{(2\pi)^4}
    \frac{(2k+q)^\mu}{k^2-m^2}  \frac{ (2k+q)^\nu}{(p+k)^2-m^2}
\ee
In the formalism of \cite{brax2018bounding} the potential is given by integrals over discontinuities of the momentum space potential $\tilde V$, 
\be
V(r)=\frac{i}{(2\pi)^2r}\int^\infty_{2m} d\lambda\lambda [\tilde V] e^{-\lambda r}\,.
\label{eq:V_int}
\ee
In the present case  the discontinuities are found to be
\be
[\tilde V^{1,{\rm massive}}_b] = \eta \frac{1}{16 \pi^2\Lambda^4}\left( 10\lambda^2 [f_1]- (8m^2+5\lambda^2)[f_0] \right) \,,
\ee
\be
[\tilde V^{1,{\rm massless}}_b] = \eta \frac{1}{16 \pi^2\Lambda^4}\left( \frac{21}{2}\lambda^2 [f_1]- \Big(\frac{17}{2}m^2+5\lambda^2 \Big)[f_0] \right)\,,
\ee
\be
[\tilde V^{1,{\rm massive}}_{b,J}] = \eta \frac{1}{16 \pi^2\Lambda^4}\left( 10\lambda^2 [f_1]- (8m^2+\lambda^2)[f_0] \right) \,.
\ee
Integration of \eqref{eq:V_int} then gives respectively the results \eqref{eq:V1b_massless}, \eqref{eq:V1b_massive}, \eqref{eq:V1bJ}.

\section{Background-Induced Amplitudes}\label{app:amplitudes}

\begin{align}
    i\mathcal{M}_a^0
    &=
  - i   \frac{2^{\eta-1}}{\Lambda^2}(\bar{u}_{p_1'} u_{p_1}\bar{u}_{p_2'}  u_{p_2})\int \frac{d^4k}{(2\pi)^4} 2\pi \delta(k^2-m^2)\Big(\theta(k^0)f_+({\bm k}) + \theta(-k^0)f_-({\bm k})\Big)\times \nonumber \\&\Bigg[\frac{1}{(k + q)^2 - m^2 + i\epsilon} + \frac{1}{(k - q)^2 - m^2 + i\epsilon} \Bigg]
  \label{eq:amplitude_a0}
\\
i\mathcal{M}_b^0
    &=
 -i   \frac{\eta}{\Lambda^4}(\bar{u}_{p_1'}\gamma^\mu  u_{p_1}\bar{u}_{p_2'}\gamma^\nu  u_{p_2})\int \frac{d^4k}{(2\pi)^4} 2\pi \delta(k^2-m^2)\Big(\theta(k^0)f_{+}(\boldsymbol{k}) + \theta(-k^0)f_{-}(\boldsymbol{k}) \Big) \times \nonumber \\ &\left[\frac{(2k_\mu+q_\mu)(2k_\nu+q_\nu)}{(k + q)^2 - m^2 + i\epsilon} + \frac{(2k_\mu-q_\mu)(2k_\nu-q_\nu)}{(k - q)^2 - m^2 + i\epsilon} \right]
 \\
    i\mathcal{M}_c^0
    &=
 -i   \frac{2^{\eta-1}}{\Lambda^6}(\bar{u}_{p_1'}  u_{p_1}\bar{u}_{p_2'} u_{p_2})\int \frac{d^4k}{(2\pi)^4} 2\pi \delta(k^2-m^2)\Big(\theta(k^0)f_{+}(\boldsymbol{k}) + \theta(-k^0)f_{-}(\boldsymbol{k}) \Big) \times \nonumber \\ &\left[\frac{(k^2 + k\cdot q)^2}{(k + q)^2 - m^2 + i\epsilon} + \frac{(k^2 - k\cdot q)^2}{(k - q)^2 - m^2 + i\epsilon} \right]  
\\
    i\mathcal{M}_a^\frac{1}{2}
    &= -i
    \frac{2^{\eta-1}}{\Lambda^4}(\bar{u}_{p_1'} u_{p_1}\bar{u}_{p_2'} u_{p_2})\int \frac{d^4k}{(2\pi)^4}2\pi \delta(k^2-m^2)\Big(\theta(k^0)f_{+}(\boldsymbol{k}) + \theta(-k^0)f_{-}(\boldsymbol{k}) \Big) \times \nonumber \\ &\mathrm{Tr}\left[(\slashed{k}+m)\frac{(\slashed{k} + \slashed{q}+m)}{(k + q)^2 - m^2 +i\epsilon} + (\slashed{k}+m)\frac{(\slashed{k} - \slashed{q}+m)}{(k - q)^2 - m^2 + i\epsilon}\right]
\\
    i\mathcal{M}_b^\frac{1}{2}
    &=-i
    \frac{\eta}{\Lambda^4}(\bar{u}_{p_1'}\gamma^\mu u_{p_1}\bar{u}_{p_2'}\gamma^\nu u_{p_2})\int \frac{d^4k}{(2\pi)^4} (2\pi) \delta(k^2 - m^2) \Big(\theta(k^0)f_{+}(\boldsymbol{k}) + \theta(-k^0)f_{-}(\boldsymbol{k}) \Big)\times \nonumber \\ &\left[\frac{\mathrm{Tr}\left[(\slashed{k}+m)\gamma_\mu ( \slashed{k} + \slashed{q}+m)\gamma_\nu \right] }{(k + q)^2 - m^2 + i\epsilon} + \frac{\mathrm{Tr}\left[(\slashed{k} - \slashed{q}+m)\gamma_\mu (\slashed{k}+m)\gamma_\nu \right] }{(k - q)^2 - m^2 + i\epsilon} \right] \\
    i\mathcal{M}_c^\frac{1}{2}
    &=-i
    \frac{2^{\eta-1}}{\Lambda^4}(\bar{u}_{p_1'}\gamma^\mu u_{p_1}\bar{u}_{p_2'}\gamma^\nu u_{p_2})\int \frac{d^4k}{(2\pi)^4} (2\pi) \delta(k^2 - m^2) \left[\theta(k^0)f_{+}(\boldsymbol{k}) + \theta(-k^0)f_{-}(\boldsymbol{k}) \right])\times \nonumber \\ &\left[\frac{\mathrm{Tr}\left[(\slashed{k}+m)\gamma_\mu \gamma^{5}( \slashed{k} + \slashed{q}+m)\gamma_\nu \gamma^{5} \right] }{(k + q)^2 - m^2 + i\epsilon} + \frac{\mathrm{Tr}\left[(\slashed{k} - \slashed{q}+m)\gamma_\mu \gamma^{5}(\slashed{k}+m)\gamma_\nu \gamma^{5}\right] }{(k - q)^2 - m^2 + i\epsilon} \right]
    \\
     i\mathcal{M}_a^1 
  &=-i
  \frac{2^{\eta-1}m^4}{\Lambda^6}(\bar{u}_{p_1'} u_{p_1}\bar{u}_{p_2'} u_{p_2})\int \frac{d^4k}{(2\pi)^4} 
  2\pi \delta(k^2-m^2)\Big(\theta(k^0)f_+({\bm k}) + \theta(-k^0)f_-(\bm k)\Big) \times \nonumber \\
  &\left(g_{\mu\nu}-\frac{k_\mu k_\nu}{m^2}\right) \left[\frac{g^{\mu\nu}-\frac{(k+q)^\mu (k+q)^\nu}{m^2}}{(k + q)^2 - m^2 + i\epsilon} + \frac{g^{\mu\nu}-\frac{(k-q)^\mu (k-q)^\nu}{m^2}}{(k - q)^2 - m^2 + i\epsilon} \right]
  \\
   i\mathcal{M}_b^1
  &= - i \frac{\eta}{\Lambda^4}(\bar{u}_{p_1'}\gamma^\mu u_{p_1}\bar{u}_{p_2'}\gamma^\nu u_{p_2})\int \frac{d^4k}{(2\pi)^4}  2\pi \delta(k^2-m^2) \Big(\theta(k^0)f_+({\bm k}) + \theta(-k^0)f_-({\bm k}) \Big)\times \nonumber \\&
\Bigg[\frac{(g_{\mu\beta}(k+2q)_\alpha+g_{\mu\alpha}(k-q)_\beta  +g_{\alpha\beta}(-2k-q)_\mu)(g_{\nu\beta}(k+2q)_\alpha+g_{\nu\alpha}(k-q)_\beta  +g_{\alpha\beta}(-2k-q)_\nu)}{(k+q)^2-m^2 + i\epsilon}+ \nonumber \\
&\frac{(g_{\mu\beta}(k+q)_\alpha+g_{\mu\alpha}(k-2q)_\beta  +g_{\alpha\beta}(-2k+q)_\mu)(g_{\nu\beta}(k+q)_\alpha+g_{\nu\alpha}(k-2q)_\beta  +g_{\alpha\beta}(-2k+q)_\nu)}{(k-q)^2-m^2 + i\epsilon} \Bigg]
\end{align}
\begin{align}
  i\mathcal{M}_c^1 
  &=
- i \frac{2^{\eta+3}}{\Lambda^6}(\bar{u}_{p_1'} u_{p_1}\bar{u}_{p_2'} u_{p_2})\int \frac{d^4k}{(2\pi)^4} 2\pi \delta(k^2-m^2) \Big(\theta(k^0)f_+({\bm k}) + \theta(-k^0)f_-({\bm k}) \Big)\times \nonumber \\ &\left[\frac{2(k\cdot(k+q))^2+k^2(k+q)^2}{(k + q)^2 - m^2 + i\epsilon} + \frac{2(k\cdot(k-q))^2+k^2(k-q)^2}{(k - q)^2 - m^2 + i\epsilon}  \right]
\\  
  i\mathcal{M}_d^1 
  &=
-i  \frac{2^{\eta+4}}{\Lambda^6}(\bar{u}_{p_1'} u_{p_1}\bar{u}_{p_2'} u_{p_2})\int \frac{d^4k}{(2\pi)^4} 2\pi \delta(k^2-m^2) \Big(\theta(k^0)f_+({\bm k}) + \theta(-k^0)f_-({\bm k}) \Big)\times \nonumber \\
&\left[\frac{(k\cdot(k+q))^2-k^2(k+q)^2}{(k + q)^2 - m^2 +i\epsilon} + \frac{(k\cdot(k-q))^2-k^2(k-q)^2}{(k - q)^2 - m^2 +i\epsilon} \right] 
        \end{align}

\section{Exact Background-Induced Potentials}\label{app:potentials}

\begin{align}
    V_a^0(r) &= -\frac{2^{\eta - 1}}{16 \pi^3 \Lambda^2 r^2} \int_0^{\infty} dk \frac{k (f_+(k) + f_-(k) )}{\sqrt{k^2 + m^2}} \sin{(2kr)} \\
    V_b^0(r)
    &= - \frac{\eta}{4 \pi^3 \Lambda^4 r^2} \int_{0}^{\infty} dk k  \sqrt{k^2 + m^2} (f_{+}(k) + f_{-}(k) ) \sin{(2 k r)} \label{eq: Vb0} \\
    V_{c}^{0}(r) &= -\frac{2^{\eta-1}}{8 \pi^3\Lambda^6 r^6}\int_{0}^{\infty} dk \frac{k (f_{+}(k) + f_{-}(k))}{\sqrt{k^2 + m^2}} \times \nonumber \\ &\Big[4kr\cos{(2kr)}\Big(r^2(2k^2 + m^2)-3 \Big)+ \sin{(2kr)} \Big(6 - 2r^2(6k^2 + m^2) + r^4(2k^2 + m^2)^2\Big) \Big] \\
    V_a^{\frac{1}{2}}(r) &= -\frac{2^{\eta-1}}{4 \pi^3 \Lambda^4 r^4} \int_{0}^{\infty} dk \frac{k (f_{+}(k) + f_{-}(k))}{\sqrt{k^2 + m^2}}  \left[(2r^2(k^2 + m^2) - 1) \sin{(2kr) + 2kr\cos{(2kr)}} \right] \\
    V_b^\frac{1}{2}(r)
    &= - 
    \frac{\eta}{4 \pi^3 \Lambda^4 r^4}\int_0^\infty dk \frac{k \left(f_{+}(k) + f_{-}(k) \right)}{\sqrt{k^2 + m^2}} \left[(1 + 2 m^2 r^2)\sin{(2kr)} - 2kr\cos{(2kr)} \right] \\
     V_c^\frac{1}{2}(r)
    &=  
    -\frac{2^{\eta-1}}{4 \pi^3 \Lambda^4 r^4}\int_0^\infty dk \frac{k \left(f_{+}(k) + f_{-}(k) \right)}{\sqrt{k^2 + m^2}} \left[\sin{(2kr)} - 2kr\cos{(2kr)} \right] \\
    V_a^1(r) 
  &= -
  \frac{2^{\eta-1}}{16 \pi^3 \Lambda^6 r^6}\int_0^\infty dk
  \frac{k (f_+(k) + f_-(k))}{\sqrt{k^2 + m^2}} \Big[4 k r(m^2 r^2 + (2 k^2 r^2 - 3))\cos{(2kr)}  + \nonumber \\
  &(m^2 r^2(-2 + (4k^2 + 3m^2)r^2) + 2 (3 - 6 k^2 r^2 + 2 k^4 r^4))\sin{(2kr)}\Big] \\
  V_b^1(r)
  &= - \frac{\eta}{16 \pi^3 \Lambda^4 r^4}\int_0^\infty dk \frac{k (f_+(k) + f_-(k))}{\sqrt{k^2 + m^2}} \Big[-16k r \cos{(2kr)} + 2(4 - 3 r^2(k^2 -2 m^2))\sin{(2kr)} \Big] \\
  V_c^1(r) 
  &= - \frac{2^{\eta+3}}{ 16 \pi^3 \Lambda^6 r^6}\int_0^\infty dk \frac{k (f_+(k) + f_-(k))}{\sqrt{k^2 + m^2}} \Big[8kr(-3 + r^2(2k^2 + m^2))\cos{(2kr)} + \nonumber \\ &(12 - 4r^2(6k^2 + m^2) + r^4(8k^4 + 8k^2 m^2 + 3m^4))\sin{(2kr)} \Big] \\
  V_d^1(r)
&= - \frac{2^{\eta+5}}{16 \pi^3 \Lambda^6 r^6}\int_0^\infty dk \frac{k (f_+(k) + f_-(k))}{\sqrt{k^2 + m^2}} \Big[2kr(-3 + r^2(2k^2 + m^2))\cos{(2kr)} + \nonumber \\
&(3 - r^2(6k^2 + m^2) + 2k^2r^4(k^2 + m^2))\sin{(2kr)} \Big]
\end{align}

\section{Computing a Potential}\label{app:example}

As an example we calculate the $V_a^0$ potential with explicit phase space distribution function.

We start with the amplitude \eqref{eq:amplitude_a0}
Using the non-relativistic limit $q \simeq (0, \boldsymbol{q})$ and the identity
\begin{equation}
    \delta(k^2 - m^2) = \frac{1}{2E_k}\left[\delta(k^0 + E_k) + \delta(k^0 - E_k) \right],
\label{eq: identity_delta}
\end{equation}
we integrate $k^0$ to obtain:
\begin{align*}
    i\mathcal{M}_a^0
    =
  i   \frac{2^{\eta-1}}{\Lambda^2}(2 m \xi^{' \dagger}\xi)_{p_1}(2 m \xi^{' \dagger}\xi)_{p_2}&\int \frac{d^3k}{(2\pi)^3} \frac{\big(f_+({\bm k}) + f_-({\bm k})\big)}{2 E_k}\Bigg[\frac{1}{\boldsymbol{q}^2 + 2\boldsymbol{k}.\boldsymbol{q} - i\epsilon} + \frac{1}{\boldsymbol{q}^2 - 2\boldsymbol{k}.\boldsymbol{q} - i\epsilon} \Bigg].    
\end{align*}

Using the definition of the non-relativistic potential \eqref{eq:Vdef} we obtain
\begin{equation}
    \tilde{V}_{a}^0(\mathbf{q}) = -\frac{2^{\eta-1}}{\Lambda^2} \int \frac{d^{3}\boldsymbol{k}}{(2\pi)^3} \frac{\big(f_+({\bm k}) + f_-({\bm k})\big)}{2E_k} \left(\frac{1}{\boldsymbol{q}^2 + 2\boldsymbol{k}.\boldsymbol{q} - i\epsilon} + \frac{1}{\boldsymbol{q}^2 - 2\boldsymbol{k}.\boldsymbol{q} - i\epsilon}\right),
\end{equation}
where $E_k = \sqrt{\boldsymbol{k}^2 + m^2}$. 
We perform a Fourier transform to find the potential as a function of the distance $\boldsymbol{r}$, 
\begin{align}
    V_{a}^0(\boldsymbol{r}) &= \int \frac{d^3 q}{(2\pi)^3}e^{i\boldsymbol{q}.\boldsymbol{r}} \tilde{V}_{a}^0(\mathbf{q}) \nonumber \\ &= -\frac{2^{\eta-1}}{\Lambda^2} \int \frac{d^{3}\boldsymbol{k}}{(2\pi)^3} \frac{\big(f_+({\bm k}) + f_-({\bm k})\big)}{2E_k} \int \frac{d^3 q}{(2\pi)^3}e^{i\boldsymbol{q}.\boldsymbol{r}} \left(\frac{1}{\boldsymbol{q}^2 + 2\boldsymbol{k}.\boldsymbol{q} - i\epsilon} + \frac{1}{\boldsymbol{q}^2 - 2\boldsymbol{k}.\boldsymbol{q} - i\epsilon}\right) \nonumber \\ &= -\frac{2^{\eta-1}}{\pi^2 \Lambda^2 r} \int \frac{d^{3}\boldsymbol{k}}{(2\pi)^3} \frac{\big(f_+({\bm k}) + f_-({\bm k})\big)}{2E_k} \int_0^\infty dq \frac{q \sin{(qr)}}{(q + 2 k \cos{\gamma} - i\epsilon)(q - 2 k \cos{\gamma} - i\epsilon)} \nonumber \\
    &= -\frac{2^{\eta-1}}{2 \pi \Lambda^2 r} \int \frac{d^{3}\boldsymbol{k}}{(2\pi)^3} \frac{\big(f_+({\bm k}) + f_-({\bm k})\big)}{2E_k}\cos{(2 k r \cos{\gamma})},
\label{eq: potencial_example}    
\end{align}
where $\gamma$ is the angle between $\boldsymbol{k}$ and $\boldsymbol{q}$. The last integral can be evaluated by closing a contour upward in the complex plane. 
Assuming an isotropic distribution, where $f(\boldsymbol{k}) = f(k)$ with $k = |\boldsymbol{k}|$, we  integrate the angular part of \eqref{eq: potencial_example}, that gives
\begin{align}
    V_{a}^0(r) &= -\frac{2^{\eta-1}}{8 \pi^3 \Lambda^2 r^2} \int_0^\infty d k \frac{k \big(f_+(k) + f_-(k)\big)}{2\sqrt{k^2 + m^2}}\sin{(2 k r)}\,.
    \label{eq: potential_example_1}
\end{align}

\subsection{Maxwell-Boltzmann Distribution}
We  assume that the background obeys the Maxwell-Boltzmann distribution $f_{\rm MB}$ defined in \eqref{eq:n_dist}. 
We assume equal distribution of particles and antiparticles, such that
\begin{equation}
    f_+(k) + f_-(k) = 2 e^{-k/T}.
\end{equation}
Using this distribution, the potential \eqref{eq: potential_example_1} becomes
\begin{align}
    V_{a}^0(r) &= -\frac{2^{\eta-1}}{8 \pi^3 \Lambda^2 r^2} \int_0^\infty d k \frac{k e^{-k/T}}{\sqrt{k^2 + m^2}}\sin{(2 k r)} \nonumber \\
    &= -\frac{2^{\eta-1} T}{8 \pi^3 \Lambda^2 r^2} \int_0^\infty d y \frac{y e^{-y}}{\sqrt{y^2 + x^2}}\sin{(2 b y)}
    \label{eq: potential_example_2}
\end{align}
where we define the quantities $y = k/T$, $b = rT$ and $x = m/T$. The integral in \eqref{eq: potential_example_2} cannot be done analytically, but we can calculate it approximately in some limits. 

For $x \ll 1$ i.e. $m\ll T$, the integral reduces to
\begin{equation}
    \int_0^{\infty} dy e^{-y} \sin{(2by)} = \frac{2b}{4b^2 + 1} = \frac{2 r T}{4 r^2 T^2 + 1}\,.
\end{equation}
The potential becomes:
\begin{equation}
    V_{a}^0(r)|_{m\ll T} \simeq -\frac{2^{\eta -1} T^2}{4 \pi^3 \Lambda^2 r (4 r^2 T^2 + 1)}\,.
\end{equation}
The potential in the limit of low and high temperature is then 
\be
   V_{a}^0(r)|_{m\ll T, r\ll T^{-1}} \simeq - \frac{2^{\eta-1}T^2}{4 \pi^3 \Lambda^2 r}\,, \quad \quad 
       V_{a}^0(r)|_{m\ll T, r\gg T^{-1}} \simeq - \frac{2^{\eta-1}}{16 \pi^3 \Lambda^2 r^3}\,.
\ee

In the case $x \gg 1$, i.e. $m \gg T$, 
due to the exponential factor 
the  integrand is unsuppressed in a region for which  $y \ll x$. The integral in \ref{eq: potential_example_2}  reduces to
\begin{equation}
    \frac{1}{x}\int_0^{\infty} dy y e^{-y} \sin{(2by)} = \frac{4b}{x(4b^2 + 1)^2} = \frac{4 r T^2}{m(4r^2T^2 + 1)^2}\,.
\end{equation}
The potential becomes:
\begin{equation}
     V_{a}^0(r)|_{m\gg T} \simeq -\frac{2^{\eta-1} T^3}{2 \pi^3 \Lambda^2 m r (4 r^2 T^2 + 1)^2}\,.
\end{equation}
The potential at low and high temperature is 
\be
    V_{a}^0(r)|_{m\gg T, r\ll T^{-1}} \simeq - \frac{2^{\eta-1} T^3}{2 \pi^3 \Lambda^2 m r}\,, \quad \quad
        V_{a}^0(r)|_{m\gg T, r\gg T^{-1}} \simeq - \frac{2^{\eta-1}}{32 \pi^3 \Lambda^2 m T r^5} \,. 
\ee

\subsection{Bose-Einstein Distribution}

We similarly compute the limits assuming the Bose-Einstein distribution.  
We find
\begin{align}
    V_{a}^0(r) &= -\frac{2^{\eta-1}}{8 \pi^3 \Lambda^2 r^2} \int_0^\infty d k \frac{k}{(e^{k/T} - 1)\sqrt{k^2 + m^2}}\sin{(2 k r)} \nonumber \\
    &= -\frac{2^{\eta-1} T}{8 \pi^3 \Lambda^2 r^2} \int_0^\infty d y \frac{y}{(e^y - 1)\sqrt{y^2 + x^2}}\sin{(2 b y)}\,.
    \label{eq: potential_example_3}
\end{align}

In the $x\ll 1$ limit the integral reduces to
\begin{equation}
    \int_0^{\infty} dy \frac{\sin{(2by)}}{e^{y} - 1} = \frac{2\pi b \coth{(2\pi b) - 1}}{4b} = \frac{2\pi r T \coth{(2\pi r T) - 1}}{4rT}\,.
\end{equation}
The resulting background-induced potential is
\begin{equation}
    V_{a}^0(r)|_{m\ll T} \simeq \frac{2^{\eta-1}(1 - 2\pi r T \coth{(2\pi r T)})}{32 \pi^3 \Lambda^2 r^3}\,.
\end{equation}

The potential at low and high temperature is 
\be
     V_{a}^0(r)|_{m\ll T, r \ll T^{-1}} \simeq -\frac{2^{\eta-1}T^2}{24\pi \Lambda^2 r}\,,\quad \quad
         V_{a}^0(r)|_{m\ll T, r \gg T^{-1}} \simeq - \frac{2^{\eta-1} T}{16\pi^2 \Lambda^2 r^2}\,    
\ee
where we used $ x \coth{x} \simeq 1 + x^2/3$.

In the case $x \gg 1$, the integral in \eqref{eq: potential_example_3} is approximately
\begin{align}
    \frac{1}{x} \int_0^{\infty} dy \frac{y \sin{(2by)}}{e^{y} - 1} &= -\frac{i}{2x} \left[\psi_1(1 - 2ib) - \psi_1(2ib + 1) \right] \nonumber \\ &= -\frac{iT}{2m} \left[\psi_1(1 - 2irT) - \psi_1(2irT + 1) \right],
\end{align}
where $\psi_n(z)$ is the polygamma function. The background-induced potential becomes
\begin{equation}
    V_{a}^0(r)|_{m\gg T} \simeq \frac{2^{\eta-1} T^2}{16 \pi^3 \Lambda^2 m r^2} i\left[\psi_1(1 - 2irT) - \psi_1(2irT + 1) \right]\,.
\end{equation}

To find the limits, we use the expansions
\begin{align}
    i\left[\psi_1(1 - 2irT) - \psi_1(2irT + 1) \right] &\simeq -8 r T \zeta(3) + \ldots \: \: \: (r T \ll 1) \\
    i\left[\psi_1(1 - 2irT) - \psi_1(2irT + 1) \right] &\simeq -\frac{1}{r T} + \ldots \: \:\: \: \: \: \: \: \: \: \: \: (r T \gg 1)\,,
\end{align}
where $\zeta(s)$ is the Riemann zeta function.
The potential at low and high temperature is
\be
        V_{a}^0(r)|_{m\gg T,r \ll T^{-1}} \simeq -\frac{2^{\eta-1} \zeta(3) T^3}{2 \pi^3 \Lambda^2 m r}\,, \quad\quad
            V_{a}^0(r)|_{m\gg T,r \gg T^{-1}} \simeq -\frac{2^{\eta-1}T}{16 \pi^3 \Lambda^2 m r^3}\,.
\ee

\subsection{Dark Matter Distribution}
Using the dark matter distribution \eqref{eq:DM_distribution},
\begin{align}
    V_a^0(r) &= -\frac{2^{\eta - 1}}{8 \pi^3 \Lambda^2 r^2} \frac{1}{N_{\mathrm{esc}}} \left(\frac{2 \sqrt{\pi}}{m v_0} \right)^3 \frac{n_0}{g} \int_0^{m v_{\mathrm{esc}}} dk \frac{k}{\sqrt{k^2 + m^2}}e^{- \frac{k^2}{m^2 v_0^2}}\sin{(2 k r)} \nonumber \\
    &= -\frac{2^{\eta - 1}}{8 \pi^3 \Lambda^2 r^2} \frac{1}{N_{\mathrm{esc}}} \left(\frac{2 \sqrt{\pi}}{m v_0} \right)^3 \frac{n_0}{g} (m v_0) \int_0^{v_{\mathrm{esc}}/v_0} dy \frac{y}{\sqrt{y^2 + x^2}}e^{-y^2}\sin{(2 b y)},\label{eq: integral_DM}
\end{align}
where we make the substitutions $y=k/(m v_0)$, $b = r m v_0$ and $x = 1/v_0$.  We apply some approximations to solve this integral. Since $v_0$ is non-relativistic ($v_0 \ll 1$, $x \gg 1$) the exponential suppression implies that the integrand is significant only in the region where $y \ll x$. Additionally, the function $y e^{-y^2}$ rapidly approaches zero for $y > 1$. Given that $v_{\mathrm{esc}}/v_0 > 1$, we can extend the upper limit of integration to infinity. Thus, the integral in \eqref{eq: integral_DM} can be approximated as  a Gaussian integral
\begin{equation}
    \frac{1}{x} \int_0^{\infty} dy y e^{-y^2}\sin{(2 b y)} = \frac{\sqrt{\pi}}{2 x} b e^{- b^2} = \frac{\sqrt{\pi}}{2} r m v_0^2 e^{-r^2 m^2 v_0^2}.
\end{equation}

The resulting background-induced potential is
\begin{equation}
    V_a^0(r) = - \frac{n_0}{4 \pi \Lambda^3 m r} e^{- r^2 m^2 v_0^2},
\end{equation}
where, in our approximation, the normalization factor becomes $N_{\mathrm{esc}} = 1$.
The potential at short and long distance are
\be
        V_{a}^0(r)|_{r \ll (m v_0)^{-1}} \simeq - \frac{n_0}{4 \pi \Lambda^3 m r}\,, \quad\quad
            V_{a}^0(r)|_{r \gg (m v_0)^{-1}} \simeq - \frac{n_0}{4 \pi \Lambda^3 m r} e^{- r^2 m^2 v_0^2}\,.
\ee

\bibliographystyle{JHEP}
\bibliography{biblio}

\end{document}